\documentclass[lettersize,journal]{IEEEtran}
\usepackage{amsmath,amssymb,amsfonts}
\usepackage{algorithmic}
\usepackage{algorithm}
\usepackage{array}
\usepackage[caption=false,font=normalsize,labelfont=sf,textfont=sf]{subfig}
\usepackage{textcomp}
\usepackage{stfloats}
\usepackage{url}
\usepackage{verbatim}
\usepackage{graphicx}
\usepackage{cite}
\usepackage{multirow}
\usepackage{xcolor}
\usepackage{arydshln}
\usepackage {booktabs}
\definecolor{EMgray}{gray}{0.45}

\usepackage{hyperref}

\def\BibTeX{{\rm B\kern-.05em{\sc i\kern-.025em b}\kern-.08em
    T\kern-.1667em\lower.7ex\hbox{E}\kern-.125emX}}

\def\orcid#1{\kern .08em\href{https://orcid.org/#1}{\includegraphics[bb=0 0 128 128,keepaspectratio,width=0.8em]{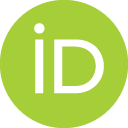}}}

\begin{document}

\title{M2D-CLAP: Exploring General-purpose Audio-Language Representations Beyond CLAP}

\author{Daisuke~Niizumi\orcid{0000-0002-5063-0508}, Daiki~Takeuchi, Masahiro Yasuda, Binh Thien Nguyen, Yasunori~Ohishi\orcid{0000-0002-7856-248X},~\IEEEmembership{Member,~IEEE,} and Noboru~Harada\orcid{0000-0002-1759-4533},~\IEEEmembership{Senior~Member,~IEEE}
        % <-this % stops a space
\thanks{Manuscript received December 21, 2021; revised September 18, 2022.}%
\thanks{The authors are with NTT Communication Science Laboratories, Nippon Telegraph and Telephone Corporation,  Atsugi 243-0198, Japan (e-mail: daisuke.niizumi@ntt.com; d.takeuchi@ntt.com; masahiro.yasuda@ntt.com; binhthien.nguyen@ntt.com; yasunori.ohishi@ntt.com; harada.noboru@ntt .com)}
}

% The paper headers
\markboth{Journal of \LaTeX\ Class Files,~Vol.~14, No.~8, August~2021}%
{Shell \MakeLowercase{\textit{et al.}}: A Sample Article Using IEEEtran.cls for IEEE Journals}

\IEEEpubid{0000--0000/00\$00.00~\copyright~2021 IEEE}

\maketitle

\begin{abstract}
Contrastive language-audio pre-training (CLAP), which learns audio-language representations by aligning audio and text in a common feature space, has become popular for solving audio tasks.
However, CLAP's audio features lack generalizability, whereas self-supervised learning (SSL) models offer general-purpose features that perform well across diverse audio tasks.
We aim to develop a broadly applicable audio representation and hypothesize that a model that learns both general audio and CLAP features should achieve our goal, which we call a general-purpose audio-language representation.
To implement our hypothesis, we propose M2D-CLAP, the first approach to jointly learn effective general audio and CLAP features. It extends an SSL masked modeling duo (M2D) by incorporating CLAP and utilizes LLM-based sentence embeddings.
The training process consists of multiple stages. In the first stage, generalizable audio features are pre-trained via a multitask objective combining M2D and CLAP, with CLAP leveraging LLM-based semantic embeddings to distill semantic knowledge into them. In the following stages, CLAP features are pre-trained and refined with guidance from the learned audio features.
Experiments demonstrated that M2D-CLAP learns high-performing general audio features (e.g., AudioSet mAP of 49.0, SOTA results in music tasks) and CLAP features, thereby enabling a general-purpose audio-language representation.\end{abstract}


\section{Introduction}
Contrastive language-image pre-training (CLIP)\cite{CLIP} has enabled the alignment of two modalities and has significantly impacted diverse domains. In the audio domain, contrastive language-audio pre-training (CLAP) has extended CLIP to audio\cite{AudioCLIP}\cite{Wav2CLIP}\cite{LAION-CLAP}\cite{CLAP2022}, followed by various advancements\cite{CLAP2023}\cite{Mei2023WavCaps}\cite{FLAP}\cite{silva2023collat}\cite{Zhu24Cacophony}\cite{niizumi24M2D-CLAP} that have exhibited superior performance for applications such as zero-shot classification and audio captioning\cite{Mei2023WavCaps}\cite{EnCLAP}\cite{Salewski2023ZeroShotAudio}.
CLAP can map audio and captions into features aligned in a common feature space; therefore, these features can be effective for various downstream application tasks since many task settings are typically related to language, such as inference of class labels.

While the CLAP models have exhibited versatility in audio-language tasks, their audio features do not generalize well in conventional audio tasks because they typically use audio encoders based on supervised learning models.
The state-of-the-art (SOTA) audio representations are self-supervised learning (SSL) models known to generalize to diverse downstream tasks\cite{LIU2022ASurvey}. In contrast, the supervised learning models, and thus the CLAP models, suffer on tasks different from the pre-training, as we reported in \cite{niizumi24M2D-CLAP}\cite{niizumi2023byol-a}.
For example, they perform well in ESC-50\cite{piczak2015esc50}, an environmental sound classification similar to the pre-trained AudioSet\cite{gemmeke2017audioset} but poorly in VoxCeleb1\cite{voxceleb}, a speaker identification task.
However, some studies have applied CLAP audio features to tasks where SSL audio features may perform well, such as \cite{DeepfakeByMSCLAP}\cite{RandByMSCLAP}.

We pursue representation learning that can be widely effective in diverse audio applications and hypothesize that a generally useful method would be one that learns both general-purpose audio features and CLAP features; we call it a \textit{general-purpose audio-language representation}.
Since the current mainstream audio representation learning methods are based on SSL with a masked prediction objective, we think combining CLAP and masked prediction-based SSL should achieve our goal.
However, as contrastive learning in CLAP is memory-intensive, naively combining it with SSL requires large computational resources.

\IEEEpubidadjcol

On the other hand, to learn the best of CLAP, a SOTA sentence embedding model based on a large-language model (LLM) would be a good candidate as a CLAP text encoder.
In the natural language processing (NLP) domain, sentence embedding models have recently progressed to exploit an LLM\cite{NV-Embed}.
LLMs have demonstrated overwhelming ability in language tasks\cite{Mistral7B} and have the potential to map audio captions into better semantic features based on knowledge accumulated in huge parameters learned from massive data.

\begin{figure}[tbp]
  \centering
  \includegraphics[width=\columnwidth]{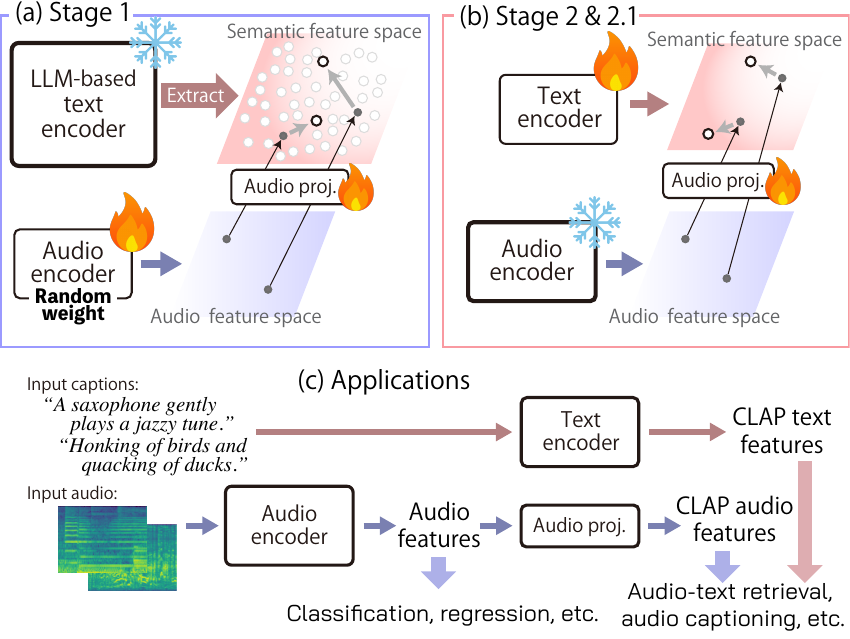}
  \vspace{-10pt}
  \caption{M2D-CLAP overview with multi-stage pre-training strategy. (a) The first stage pre-trains the audio encoder/projector from scratch to learn audio features that align with LLM-based text sentence embeddings extracted in advance; (b)~the following stages pre-train the text encoder/audio projector to learn CLAP features that align better in the semantic feature space. (c) We transfer the learned audio encoder for encoding general-purpose audio features and the learned audio projector/text encoder for encoding CLAP features.}
  \label{fig:overview}
  \vspace{-10pt}
\end{figure}

To implement a general-purpose audio-language representation and achieve its highest potential, we propose M2D-CLAP, the first approach to simultaneously achieve effective general audio and CLAP features. It extends a masked prediction-based SSL, M2D\cite{M2D2024TASLP}, combines it with a CLAP objective, and exploits the strong LLM-based sentence embeddings, as shown in Fig.~\ref{fig:overview}.

M2D-CLAP adopts a multi-stage pre-training strategy that leverages LLM-based sentence embeddings to learn high-quality features while avoiding substantial computational costs.
In the first stage, M2D-CLAP performs multitask learning with M2D and CLAP objectives to acquire general-purpose audio features aligned with LLM-based sentence embeddings, aiming to distill LLM knowledge into audio representations. These features are then fine-tuned on AudioSet to further enhance their effectiveness, particularly for CLAP.
In the subsequent stages, M2D-CLAP pre-trains a smaller non-LLM text encoder to learn CLAP text features that align with the learned audio features, aiming to transfer the knowledge into the text features.
By training either the audio or text encoder at each stage, M2D-CLAP achieves efficient training without incurring the high computational cost.

The experiments validate the effectiveness of M2D-CLAP in achieving our goal of learning a general-purpose audio-language representation. The resulting audio and CLAP features show top performance both as frozen and fine-tuned audio features, as well as in audio-language tasks. 
For a successful pre-training by multitasking M2D and CLAP, we find it essential to balance the loss between M2D and CLAP to appropriately control the strong semantic supervision. Additionally, the audio projector capacity, the use of an LLM-based text encoder, and the multi-stage training strategy all prove to be critical design choices, as demonstrated in the ablation studies in Section \ref{sec:exp-ablations}.

Our contributions are:
\begin{itemize}
\item Introducing a general-purpose audio-language representation that provides general audio and CLAP features to serve a wide range of audio and audio-language tasks.
\item Proposing M2D-CLAP, the first (to our knowledge) method to jointly learn top-performing general-purpose audio and CLAP features, thereby realizing a general-purpose audio-language representation.
\item  Providing empirical findings on combining SSL and CLAP, along with publicly available code and pre-trained weights\footnote{\scriptsize{\url{https://github.com/nttcslab/m2d}}} to support future research.
\end{itemize}

The differences from our conference paper\cite{niizumi24M2D-CLAP} are summarized in the next section.

\section{Related Work}
\subsection{Relationship with our previous work}
We introduced M2D-CLAP in our previous work\cite{niizumi24M2D-CLAP}. While it demonstrated the effectiveness, the approach left room for improvement, particularly in the performance of the CLAP features.
This paper redefines and extends M2D-CLAP to enhance the effectiveness of its audio and CLAP features, leveraging a newly introduced multi-stage pre-training strategy, a transformer-based audio projector, and LLM-based sentence embeddings (Section \ref{sec:m2d-clap}).
To further assess its effectiveness, we also conducted experiments on audio-text retrieval, audio captioning, and ablation analyses in addition to transfer learning and zero-shot classification (Sections \ref{sec:exp-audio-feature})--\ref{sec:exp-ablations}).

To make the difference from the redefined M2D-CLAP clear, we specifically denote it as M2D-CLAP$_\text{2025}$ and the previous M2D-CLAP\cite{niizumi24M2D-CLAP} as M2D-CLAP$^\text{ stage1}_\text{2024}$. Note that M2D-CLAP$^\text{ stage1}_\text{2024}$ is a variant of the first-stage M2D-CLAP.

\subsection{Conventional audio representation models}
Since the release of the large-scale dataset AudioSet\cite{gemmeke2017audioset}, various supervised learning models such as PANNs\cite{kong2020panns}, AST\cite{gong2021ast}, and HTS-AT\cite{Chen2022HTS-AT} that utilize class labels have followed. 
In contrast, SSL methods such as COLA\cite{saeed2020cola} and BYOL-A\cite{niizumi2023byol-a} are general-purpose audio representations that do not depend on task supervision, such as labels.
Masked prediction-based SSLs have progressed and demonstrated remarkable general-purpose performance. SSAST\cite{gong2022ssast}, MAE-AST\cite{Baade2022MAE-AST}, MSM-MAE\cite{niizumi2022msm-mae}, and Audio-MAE\cite{huang2022amae} learn by reconstructing the masked input, while M2D and ATST\cite{Li2023ATST-TALSP} learn by predicting the masked features. More advanced models include BEATs\cite{chen2022beats}, which learn by predicting distilled and tokenized labels, and CED\cite{dinkel2023ced}, which distills multiple large models.
While they perform well on conventional audio tasks, their features do not align with languages.

\subsection{CLAP models}
Inspired by CLIP\cite{CLIP}, the early CLAP models AudioCLIP\cite{AudioCLIP} and Wav2CLIP\cite{Wav2CLIP} learn audio features that align with the pre-trained CLIP multimodal feature. CLAP\cite{CLAP2022}\cite{CLAP2023} and LAION-CLAP\cite{LAION-CLAP} adapt CLIP to audio, WavCaps\cite{Mei2023WavCaps} improves performance with their dataset, and MGA-CLAP\cite{MGA-CLAP} further improves fine-grained audio-language alignment.
While they can provide an audio feature as a part of CLAP features, they do not generalize well on conventional audio tasks, as reported in \cite{niizumi24M2D-CLAP}\cite{niizumi2023byol-a}.

Methods that take a similar approach to M2D-CLAP are SLIP\cite{Mu2022SLIP} in the image domain (combines SSL and CLIP), FLAP\cite{FLAP} (combines MAE\cite{he2022masked} and CLAP), SupMAM-CLAP\cite{xin23d_mamclap} (combines Audio-MAE and distills a CLAP model), Cacophony\cite{Zhu24Cacophony} (combines Audio-MAE, CLAP, and a captioning model in a two-stage training), and CoLLAT\cite{silva2023collat} (CLAP with frozen text encoders).
Unlike these approaches, we combine M2D, CLAP, and the recent LLM-based sentence embeddings in a multi-stage pre-training to learn a general-purpose audio-language representation that provides effective audio and CLAP features.

\section{Proposed Method}
We propose M2D-CLAP, which extends M2D to combine it with CLAP to learn general-purpose audio-language representations that provide useful features for a broad range of conventional audio and audio-language tasks.

\subsection{Background: Masked Modeling Duo}
M2D is an SSL framework applicable to 2D structured data input such as audio spectrograms, and it pre-trains Vision Transformer\cite{ViT} (ViT) with a masked prediction objective to learn a general-purpose representation.

\subsubsection{Pre-training}
As shown in Fig. \ref{fig:system}(a), M2D consists of two networks, the online and the target, and learns representations by predicting the target output features using the online output features.
It takes a spectrogram with a predefined size (e.g., 80 frequency bins and 608 time steps) as the input, which is split into patches (e.g., $16\times 16$), mapped by a learnable linear layer (patch embedding layer), and handled as a series $x$ (e.g., $(80/16)\times(608/16)=190$ patches).
M2D then adds positional encoding to patches and randomly selects a number of patches according to a masking ratio as masked patches $x_m$ (e.g., 70\% of the input patches) and the rest as visible patches $x_v$ (e.g., the remaining 30\%).

The online network with a set of weights $\theta$ encodes $x_v$ using the online audio encoder $f_\theta$ into the audio features $z_v = f_\theta(x_v)$.
It concatenates the learnable mask tokens $m$ to $z_v$, adds the position encoding $p$, and inputs them to the predictor $g_\theta$ to predict the target features as $\hat{z} = g_\theta(\text{concat}(z_v, m) + p)$.
It then outputs the prediction result $\hat{z}_m = \{\, \hat{z}[i] \mid i \in I_M \,\}$ of the masked patch features, where $I_M$ is the set of masked patch indices.

The target network, defined by parameter $\xi$, outputs the features $z_m = f_\xi(x_m)$ and standardizes them to the final target output $\tilde{z}_m = ({z_m - \text{mean}{(z_m)}})/{\sqrt{\text{var}{(z_m)}}}$.
The loss is calculated using the online prediction $\hat{z}_m$ against the actual target output $\tilde{z}_m$ as a training signal by the mean square error (MSE) of $l_2$-normalized $\hat{z}_m$ and $\tilde{z}_m$:
\vspace{-0.1cm}
\begin{equation}
L_\text{m2d} \triangleq ||l_2(\hat{z}_m) - l_2(\tilde{z}_m)||^2_2 = 2 - 2 \cdot \frac{\langle \hat{z}_m, \tilde{z}_m \rangle }{||\hat{z}_m||_2 \cdot ||\tilde{z}_m||_2},
\label{eq:eq-byol-mse}
\vspace{-0.1cm}
\end{equation}
where $\langle\cdot, \cdot\rangle$ denotes the inner product.

The M2D framework updates $\theta$ only to minimize the loss $L_\text{m2d}$, as depicted by the stop-gradient in Fig. \ref{fig:system}(a), and updates $\xi \leftarrow \alpha \xi + (1 - \alpha) \theta$ as an exponential moving average of $\theta$ with a decay rate $\alpha$.
M2D exploits the momentum encoder to learn effective representations from the target network.

\subsubsection{Transfer learning}
After the pre-training, only the audio encoder $f_\theta$ is transferred from the online network to the downstream tasks.
As the encoder accepts the input no longer than the predefined duration (e.g., 608 time steps) in the pre-training stage, the inference procedure handles variable-length audio inputs in two ways: one is by cutting the input with the predefined duration and repeatedly encoding the chunks; the other is by interpolating the positional encoding to the input audio duration (i.e., stretching the acceptable input duration). The latter is used for fine-tuning the encoder to a task with an input duration similar to the pre-training, and the former is used in all other cases, such as linear evaluation.

According to the task requirements, the encoder outputs a feature for each patch $z$, which is summarized into a time frame-level feature $z'$ or an audio clip-level feature $z''$. The following pseudo-code calculates $z'$ for each time frame:
\begin{equation}
\begin{split}
z' = & z.\text{reshape}(B,  N_F, N_T, D)\\
    & .\text{transpose}(1, 2)\\
    &.\text{reshape}(B, N_T, N_F D), \label{eq:msm-mae}
\end{split}    
\end{equation}
where $z=f_\theta(x) \in R^{B \times N_F N_T \times D}$, $B$ is batch size, $N_F$ is the number of patches along frequency, $N_T$ is the number of patches along time, $D$ is a feature dimension, and $z' \in R^{B \times N_T \times N_F D}$ is the calculation result.
The audio clip-level feature $z''$ is calculated by averaging $z'$ over time: $z'' = 1/N_T\sum_{t=1}^{N_T} z'[t]$.
For example, $z''$~becomes a 3840-d feature, where $D=768$ and $N_F=5$.

\subsubsection{M2D for X}
In addition to SSL, M2D provides a framework for learning a specialized representation in application \textit{X} (M2D-X) by extending M2D to add an application-specific task and form a multitask of M2D and the added task in parallel. During the pre-training, the intermediate outputs $z_v$ and $\hat{z}_m$ in M2D are concatenated and fed into the additional task, and the loss of M2D and the additional task are combined to train the entire network.
As a result, the learned representation reflects both objectives from M2D and the added task.

\subsection{Proposed M2D-CLAP: M2D Meets CLAP} \label{sec:m2d-clap}
\begin{figure}[tbp]
  \vspace{0pt}
  \centering
  \includegraphics[width=1.0\columnwidth]{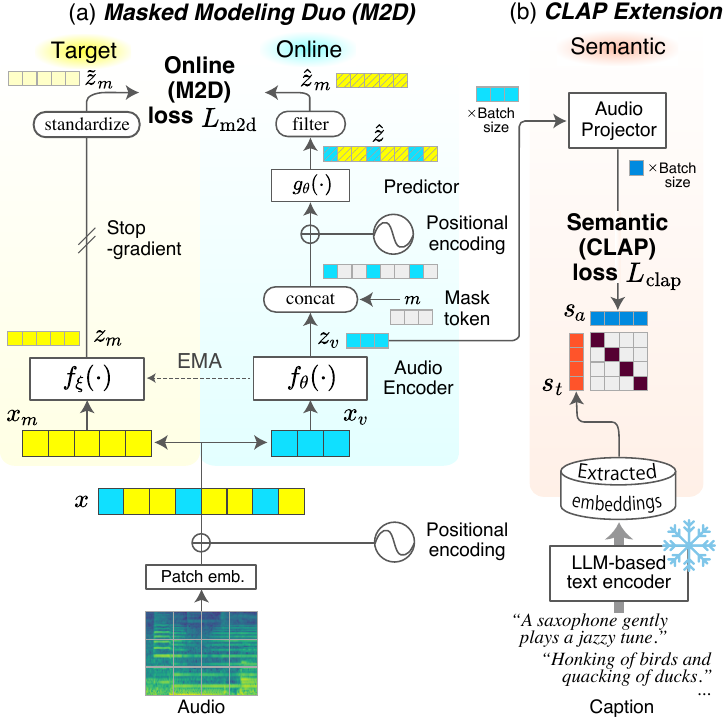}
  \vspace{-15pt}
  \caption{M2D and M2D-CLAP first stage pre-training flows: (a) M2D learns general-purpose audio features through a task to predict the masked features (yellow) using the visible features (blue). (b) M2D-CLAP extends M2D with CLAP and jointly learns CLAP features using contrastive learning between audio (blue) and text (red) features. The text features are the LLM-based sentence embeddings of the audio captions extracted in advance.}
  \label{fig:system}
  \vspace{-10pt}
\end{figure}

M2D-CLAP is an M2D-X variant that combines M2D and CLAP training objectives to implement a general-purpose audio-language representation.
As shown in Fig. \ref{fig:system}, it takes a multitask approach comprising (a) M2D and (b) CLAP that learns from paired audio and caption data.
The CLAP extension has a semantic network branch that conducts contrastive learning among audio and text features to align them in a semantic feature space.

M2D-CLAP employs a multi-stage training strategy to learn the best of audio and CLAP features under a multitask of M2D and CLAP without the cost of huge computational resources. In the first stage, it learns effective audio features from M2D that also jointly learns from powerful LLM-based sentence embeddings by aligning the audio features with them through CLAP, aiming to achieve the best audio features that are also effective as CLAP audio features. In the second stage, it learns effective CLAP text features that align with the learned audio features in a semantic feature space. In each stage, it additionally fine-tunes features to achieve effective CLAP performance. Through these stages, M2D-CLAP learns effective audio and CLAP features.

\subsubsection{First stage pre-training}
In the first stage, captions are encoded in advance into embeddings using an LLM-based sentence embedding model, ensuring that the training process receives only audio inputs alongside the corresponding fixed text embeddings.
M2D then conducts its training process and feeds the audio encoder's output feature $z_v$ to the semantic network. While the original M2D-X framework provides $\hat{z}_m$ in addition to $z_v$, M2D-CLAP uses $z_v$ only for training stability.
The audio projector converts the audio features $z_v$ into semantic features $s_a$. The text embeddings are used as they are or mapped to the same dimension as $s_a$ using a learnable linear layer and handled as text semantic features $s_t$. The training process does not involve the LLM-based sentence embedding model, which requires a large GPU memory.

This stage aims to train an effective audio encoder, so it is crucial that the training signal, semantic feature $s_t$, precisely represents the semantics of the text. We use a powerful LLM-based sentence embedding model and \textit{distill} its semantic knowledge into the audio encoder through CLAP.
The audio projector is a transformer encoder to summarize $z_v$ so that the $s_a$ can better represent the semantics of an input audio.

The training process calculates the CLAP loss using the cosine similarity $S_{mn}$ between $s_a$ and $s_t$:
%\vspace{-0.3cm}
\begin{equation}
S_{mn} = \frac{\langle s_a^{(m)}, s_t^{(n)} \rangle }{||s_a^{(m)}||_2 \cdot ||s_t^{(n)}||_2},
\label{eq:eq-cossim}
\end{equation}
%\vspace{-0.2cm}
where $s_a^{(m)}$ is the semantic feature of the $m$th audio batch sample, and $s_t^{(n)}$ is the semantic feature of the $n$th caption batch sample.
The CLAP loss $L_\text{clap}$ is the average of the NT-Xent\cite{chen20simclr} losses calculated along the audio and caption axes:
%\vspace{-0.1cm}
\begin{equation}
L_\text{clap} = \text{\scalebox{0.85}{%
$-\frac{1}{2B} \sum^B_i \left[ \log{\frac{\exp{S_{ii}/\tau}}{\sum^B_j \exp{S_{ji}/\tau}}} + \log{\frac{\exp{S_{ii}/\tau}}{\sum^B_j \exp{S_{ij}/\tau}}} \right],$}}
\label{eq:eq-clip-loss}
\end{equation}
where $B$ is the number of batch samples, and $\tau$ is a learnable temperature parameter. We follow CLIP \cite{CLIP} to initialize $\tau$ with 0.07 and clip it to prevent scaling the logits by more than 100 for training stability.

The entire loss $L_\text{stage1}$ combines $L_\text{m2d}$ and $L_\text{clap}$:
%\vspace{-0.1cm}
\begin{equation}
L_\text{stage1} = \lambda_\text{m2d} L_\text{m2d} + \lambda_\text{clap} L_\text{clap},
\label{eq:eq-m2d-clap-loss}
\end{equation}
where the loss weights $\lambda_\text{m2d}$ and $\lambda_\text{clap}$ control the contribution.

The first-stage pre-training combines supervised learning (CLAP) and SSL (M2D), enabling the audio encoder to learn effective features through their synergy. CLAP directs feature learning toward the semantics of the caption text but may miss aspects of sound not explicitly described. In contrast, SSL promotes general-purpose feature learning but lacks the guidance provided by textual supervision. By integrating both, the model learns semantically aligned yet generalizable representations. 
The learned audio features improve performance on tasks such as environmental sound and music compared to M2D alone, at the cost of slightly reduced performance on speech tasks, but with an overall performance gain. The experiments in Section \ref{sec:exp-audio-feature} demonstrate this trade-off.

\paragraph{Additional stage 1.1: AudioSet fine-tuning} \label{sec:additional-stage-1.1}
This stage fine-tunes the first-stage pre-trained audio encoder to AudioSet under a supervised setting using labels. 
This additional process corresponds to adopting the AudioSet fine-tuned audio encoders in the conventional CLAP models\cite{LAION-CLAP}\cite{CLAP2022}\cite{CLAP2023}.
Experiments in Sections \ref{sec:exp-clap-atr} and \ref{sec:exp-clap-captioning} show that this fine-tuning step is advantageous with AudioCaps\cite{kim2019audiocaps} (a subset of AudioSet).

\begin{figure}[tbp]
  \vspace{0pt}
  \centering
  \includegraphics[width=0.73\columnwidth]{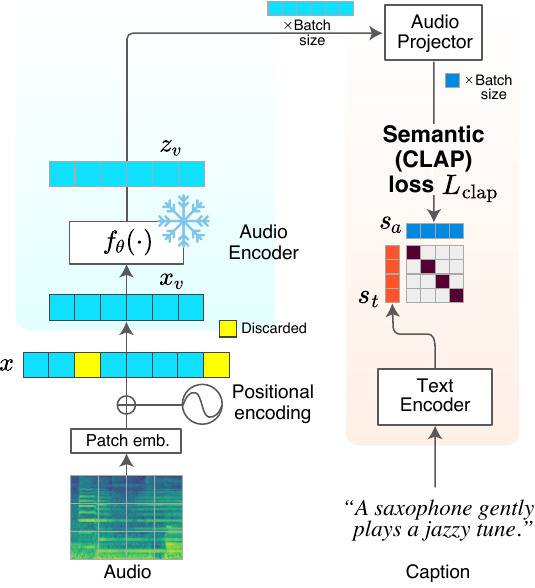}
  \vspace{-5pt}
  \caption{M2D-CLAP second stage pre-training flow: M2D-CLAP learns only from the CLAP objective. As in M2D, the audio patch embeddings are partially masked and discarded with a low masking ratio. This stage trains the text encoder and audio projector while the audio encoder is frozen.}
  \label{fig:system-2}
  \vspace{-10pt}
\end{figure}

\subsubsection{Second stage pre-training}
The second stage pre-trains the text encoder and audio projector using CLAP alone and freezes the audio encoder weights so that we preserve effective audio features learned in the first stage, as shown in Fig. \ref{fig:system-2}.
To facilitate CLAP training with a precise loss, we input the entire audio sample without truncation and apply a low masking ratio, allowing a nearly consistent loss calculation.

In this stage, we employ a modest capacity model, such as BERT, as the text encoder to maintain effective contrastive learning with a sufficient batch size.
The training process masks the audio patches with a low masking ratio (e.g., 30\%), the frozen audio encoder encodes input patches $x_v$ into the features $z_v$, and the audio projector summarizes them into the semantic features $s_a$. The text encoder encodes captions into the semantic features $s_t$, and the clap loss $L_\text{clap}$ is calculated between $s_a$ and $s_t$, which is the loss for the second stage.

\paragraph{Additional stage 2.1: WavCaps fine-tuning}
This stage further fine-tunes the second-stage audio projector and text encoder to refine the CLAP features using the same dataset as in WavCaps\cite{Mei2023WavCaps} and with no masking of the audio input to learn the best text/audio alignment. We show that this process improves performance in audio-language tasks in Section \ref{sec:eval-clap-feature}.

\subsubsection{M2D-CLAP transfer learning}
Once pre-trained, we transfer the audio encoder, audio projector, and text encoder as models constituting a general-purpose audio-language representation to the downstream tasks. The audio encoder outputs audio features $z$ of the input audio, the audio projector converts $z$ into CLAP audio features $s_a$, and the text encoder encodes captions into CLAP text features $s_t$. While the audio feature $z$ is useful for conventional audio tasks, $s_a$ and $s_t$ work as CLAP features in audio-language tasks such as zero-shot classification and audio-text retrieval.

%\subsubsection{Additional pre-training stages}
%While the two training stages enable the learning of general-purpose features, the experiments show that these features still leave a gap with top performance in existing benchmarks if they are \textit{not fine-tuned}.
%Therefore, to facilitate the comparison with previous studies, we also use two additional stages to bring the experimental conditions closer to those of existing studies.

\subsubsection{Training design and efficiency}
While other strategies are possible, we pre-train the audio encoder first to leverage LLM knowledge for both audio and text features. The first stage of M2D-CLAP enables this, unlike typical CLAP models that use an existing audio encoder pre-trained only on audio data. To train a CLAP model in practice, pre-trained audio and text encoders are required; thus, we train the audio encoder first and the text encoder in the second stage. To keep training efficient without high computational cost, we train the encoders separately and use the LLM offline.

We utilize an audio projector only, unlike previous CLAP, which typically uses both audio and text projectors. While an audio projector is crucial to bridge the gap between the descriptive general audio and abstract (semantic) CLAP features, a text projector degrades performance in M2D-CLAP, as verified in Section \ref{sec:exp-abl-training-design}.

The staged pre-training not only enables learning a well-performing representation but also allows training with reduced computational resources.
In the first stage, M2D-CLAP learns from LLM features that remain fixed for identical captions, enabling a substantial reduction in computational cost by pre-computing features for all captions. During training, these pre-computed features are used instead of loading the LLM instance, saving computational resources.
In contrastive learning, a large batch size is critical for sufficient training. While FLAP, a similar method, requires $36\times64$ GPUs to achieve a batch size of 2304, M2D-CLAP can conduct training with 4 A100 GPUs for a batch size of 2048.

\section{Experiments}\label{sec:experiments}
To assess the effectiveness of M2D-CLAP representations, we evaluated the audio features on conventional audio tasks under settings of frozen audio features (Sections \ref{sec:exp-audio-LE} and \ref{sec:exp-audio-music}) and fine-tuning audio features (Section \ref{sec:exp-audio-FT}) and the CLAP features using three tasks: zero-shot classification (Section \ref{sec:exp-clap-ZS}), audio-text retrieval (Section \ref{sec:exp-clap-atr}), and audio captioning (Section \ref{sec:exp-clap-captioning}).

We also studied how the pre-training design and parameters impact the final performance (Section \ref{sec:exp-ablations}), compared the CLAP features using visualizations (Section \ref{sec:exp-abl-feature-viz-clap}), and show how M2D-CLAP focuses on the audio patches to summarize CLAP audio features (Section \ref{sec:exp-abl-feature-viz-att}).
The following sections refer to M2D-CLAP and its intermediate-stage variants used in the ablation studies as follows:
\begin{itemize}
\item M2D-CLAP$_\text{2025}$: Final M2D-CLAP representation pre-trained in all two plus two additional stages.
\item M2D-CLAP$^\text{ stage1}_\text{2025}$: First stage pre-trained representation. Combination of pre-trained audio encoder/audio projector and off-the-shelf LLM-based sentence embedding model as a text encoder.
\item M2D-CLAP$^\text{ stage2}_\text{2025}$: Second-stage pre-trained representation, combining the first-stage audio encoder with the second-stage audio projector/text encoder. The audio encoder is not fine-tuned on AudioSet; stage 1.1 is omitted.
\item M2D-CLAP$^\text{ stage2.1}_\text{2025}$: Second stage plus stage 2.1 pre-trained representation, combining the first-stage audio encoder and the stage 2.1 pre-trained audio projector/text encoder.
\item M2D-CLAP$^\text{ stage1}_\text{2024}$: The model from the initial study\cite{niizumi24M2D-CLAP} as an ablation variant of M2D-CLAP$^\text{ stage1}_\text{2025}$ for using a non-LLM text encoder and AudioSet only for audio data.
\end{itemize}

\subsection{Experimental setup}
\subsubsection{Basic setup}
We followed M2D for the basic setup and implemented M2D-CLAP on top of the M2D codebase as a variant of M2D-X\cite{M2D2024TASLP}.
Specifically, the audio encoder is ViT-base\cite{ViT} with a 768-d feature, input audio duration of 6 s, and patch size of $16\times 16$. The audio projector is a single-block transformer encoder with a 768-d feature, a single head, a 768-d feed-forward network latent feature, and a class token. It inputs audio features encoded by the audio encoder and outputs the class token feature as a summarized CLAP audio feature.

The audio samples were preprocessed to a log-scaled mel spectrogram with a sampling frequency of 16,000 Hz, window size of 25 ms, hop size of 10 ms, and 80 mel-spaced frequency bins in the range of 50 to 8000 Hz.
We standardized spectrograms with the statistics of the first stage dataset (e.g., average of $-7.26$ and standard deviation of $4.35$).

\subsubsection{Pre-training setup}
\paragraph{The first stage}
In the first stage, we used the M2D masking ratio of 0.7, and all other settings were the same as in M2D, such as 300 epochs, 20 warm-up epochs, batch size of 2048, the base learning rate of $3.0\times 10^{-4}$, and the linearly interpolated EMA decay rate starting from 0.99995 to 0.99999 at the end.
The loss weights were $\lambda_\text{m2d}=1.0$ and $\lambda_\text{clap}=0.01$.

We used a generalist text embedding model NV-Embed-v2\cite{NV-Embed}\cite{NV-Embed2} with 4096-d features as a text encoder from Hugging Face\footnote{\scriptsize{\url{https://huggingface.co/nvidia/NV-Embed-v2}}}, built on top of an LLM, Mistral 7B\cite{Mistral7B}.
NV-Embed-v2 exhibits SOTA performance on the Massive Text Embedding Benchmark (MTEB)\cite{MTEB}.
We used a learnable linear layer to map the 4096-d output feature to 768 dimensions.

\paragraph{The second stage}
In the second stage,  we used the M2D masking ratio of 0.3, 30 epochs, five warm-up epochs, batch size of 2048, and the base learning rate of $3.0\times 10^{-6}$. 
We used BERT-base\cite{bert} as a text encoder from Hugging Face\footnote{\scriptsize{\url{https://huggingface.co/google-bert/bert-base-uncased}}} with a 768-d feature dimension and used no linear layer.

\paragraph{Additional stages}
In the additional stage 1.1, we fine-tuned the first-stage audio encoder in the same way as in the fine-tuning experiments in Section \ref{sec:exp-audio-FT}. In the additional stage 2.1, we used the same setting as in the second stage once again, except for a masking ratio of 0.0 (i.e., no masking) and a limited dataset.

\subsubsection{Pre-training datasets}
\paragraph{The first stage}
In the first stage, the pre-training data consists of AudioSet\cite{gemmeke2017audioset}, VGGSound\cite{Chen20VGGSound}, and WavCaps\cite{Mei2023WavCaps}, 2,532,215 samples in total. We made it closer to CLAP$_\text{2023}$\cite{CLAP2023}; they used 4.6 M samples from AudioSet, WavCaps, and many other datasets.
We used Auto-ACD\cite{sun2023autoacd} for the caption data of AudioSet and VGGSound and supplemented the missing captions with \textit{"The sound of $\langle$labels$\rangle$"} using their labels.
Since the curated dataset included data samples for the downstream tasks (AudioCaps\cite{kim2019audiocaps}, Clotho\cite{icassp20Clotho}, ESC-50\cite{piczak2015esc50}, US8K\cite{salamon2014urbansound}) used in the evaluations, we removed all of these data samples. 
For training audio, we randomly cropped 6-s segments.

\paragraph{The second stage}
For the second stage, based on the first stage dataset, we added the AudioCaps and Clotho training samples for a fair comparison with previous studies, especially WavCaps\cite{Mei2023WavCaps}. As with the first stage, we removed all data samples of ESC-50/US8K from the added AudioCaps/Clotho training samples. The total number of samples was 2,585,227.
In this stage, we used 10-s audio data as they were during the training, or we randomly cropped for 10 s if a sample was longer.

\paragraph{Additional stages}
For the additional training stages, we used AudioSet containing 2,005,132 training and 20,178 test samples with its labels for fine-tuning in stage 1.1. In stage 2.1, we used WavCaps, AudioCaps, and Clotho, 439,743 samples in total, as in the previous CLAP studies\cite{Mei2023WavCaps}\cite{Yan24bridging}.

\begin{table*}[tbp]
\caption{Details of the downstream tasks used in the evaluations.}
\label{tab:list-ds}
\centering
\vspace{-5pt}
\resizebox{0.999\textwidth}{!}{%
\begin{tabular}{rcccccccccccc}\toprule
& \multicolumn{5}{c}{Environmental sound tasks} & \multicolumn{4}{c}{Speech tasks} & \multicolumn{3}{c}{Music tasks} \\
\cmidrule(lr){2-6} \cmidrule(lr){7-10} \cmidrule(lr){11-13}  
 & AS2M\cite{gemmeke2017audioset} & AS20K\cite{gemmeke2017audioset} & ESC-50\cite{piczak2015esc50} & US8K\cite{salamon2014urbansound} & FSD50K\cite{fonseca2020fsd50k} & SPCV2\cite{speechcommandsv2} & VC1\cite{voxceleb} & VF\cite{voxceleb} & CRM-D\cite{cao2014cremad} & GTZAN\cite{gt2013gtzan} & NSynth\cite{nsynth2017} & Surge\cite{turian2021torchsynth} \\
\midrule
Evaluation protocol$^\dagger$ & FT & FT / ZS & FT / LE / ZS & LE / ZS & ZS & FT / LE & FT / LE & LE & LE / ZS & LE / ZS & LE / ZS & LE\\
 \addlinespace[0.1cm]
Training samples & 2,005,132 & 21,940 & \multirow{3}{*}{\shortstack{5 folds\\ 2000}} & \multirow{3}{*}{\shortstack{10 folds\\ 8732}} & 40,966 & 84,843 & 138,361 & 121,281 & 5155 & 443 & 289,205 & 148,896\\
% \addlinespace[0.01cm]
Validation samples & - & - & & & - &  9981 & 6904 & 26,684 & 732 & 197 & 12,678 & 17,160 \\
Test samples & 20,178 & 20,178 & & & 10,231 & 11,005 & 8251 & 28,463 & 1551 & 290 & 4096 & 17,336 \\
 \addlinespace[0.02cm]
Classes & 527 & 527 & 50 & 10 & 200 & 35 & 1251 & 6 & 6 & 10 & 11 & 88 \\
Average duration & 10.0 s & 10.0 s & 5.0 s & 4.0 s & 7.6 s & 1.0 s & 8.2 s & 5.8 s & 2.5 s & 30.0 s & 4.0 s &  4.0 s \\
\bottomrule
\addlinespace[0.05cm]
\multicolumn{12}{l}{$^{\dagger}$FT, LE, and ZS stand for fine-tuning, linear evaluation, and zero-shot classification, respectively.}\\
\end{tabular}
}
%\vspace{-10pt}
\end{table*}

\begin{table*}[htb!]
%\vspace{-15pt}
\caption{Frozen audio feature evaluation (\%) with 95\% CI on general audio tasks under a unified setting except Pengi, Cacophony, CoLLAT, and MATPAC.}
\label{tab:results-le}
\centering
\vspace{-5pt}
\resizebox{0.98\textwidth}{!}{%
\begin{tabular}{lcllllllllll} \toprule
  & \multirow{2}{*}{\shortstack[c]{Audio enc.\\fine-tuning\\on AudioSet}} & \multicolumn{2}{c}{Env. sound tasks} & \multicolumn{4}{c}{Speech tasks} & \multicolumn{3}{c}{Music tasks} \\
 \cmidrule(lr){3-4} \cmidrule(lr){5-8} \cmidrule(lr){9-11} 
Model &   &   ESC-50 &    US8K &    SPCV2 &    VC1 &     VF &    CRM-D &    GTZAN &     NSynth &      Surge & Avg.\\
\midrule

\multicolumn{10}{l}{\textit{(Previous studies: Audio models)}}  \\
\addlinespace[0.05cm]

ATST-Clip \cite{Li2023ATST-TALSP} & & 94.1{\fontsize{6pt}{6pt} \selectfont $\pm$0.6} & 85.8 $\Lsh$ & 95.1 $\Lsh$ & 72.0 $\Lsh$ & 97.6{\fontsize{6pt}{6pt} \selectfont $\pm$0.0} & 68.8{\fontsize{6pt}{6pt} \selectfont $\pm$1.3} & 78.9{\fontsize{6pt}{6pt} \selectfont $\pm$3.5} & 76.2 $\Lsh$ & 32.8{\fontsize{6pt}{6pt} \selectfont $\pm$0.0} & 77.9 \\
ATST-Frame \cite{Li2023ATST-TALSP} & & 90.9{\fontsize{6pt}{6pt} \selectfont $\pm$0.6} & 85.8 $\Lsh$ & 94.9 $\Lsh$ &\textbf{77.4} $\Lsh$ &\textbf{98.8{\fontsize{6pt}{6pt} \selectfont $\pm$0.3}}& 72.3{\fontsize{6pt}{6pt} \selectfont $\pm$0.7} & 82.9{\fontsize{6pt}{6pt} \selectfont $\pm$6.0} & 75.9 $\Lsh$ & 40.6{\fontsize{6pt}{6pt} \selectfont $\pm$0.2} & 79.9 \\
MATPAC \cite{MATPAC2025Quelennec}  & & 93.5 $\Lsh$ & 89.4 $\Lsh$ & - & - & - & - & 85.9 $\Lsh$ &74.6  $\Lsh$ & - & - \\
Cacophony \scriptsize{(stage 1)}\cite{Zhu24Cacophony} $\dagger$ & & 87.0 $\Lsh$ & - & 92.2 $\Lsh$ & - & - & 69.7 $\Lsh$ & 83.8 $\Lsh$ & - & - & - \\
BEATs$_\text{iter3}$ \cite{chen2022beats} &  & 86.9{\fontsize{6pt}{6pt} \selectfont $\pm$1.4} & 84.8{\fontsize{6pt}{6pt} \selectfont $\pm$0.1} & 89.4{\fontsize{6pt}{6pt} \selectfont $\pm$0.1} & 41.4{\fontsize{6pt}{6pt} \selectfont $\pm$0.7} & 94.1{\fontsize{6pt}{6pt} \selectfont $\pm$0.3} & 64.7{\fontsize{6pt}{6pt} \selectfont $\pm$0.8} & 72.6{\fontsize{6pt}{6pt} \selectfont $\pm$4.3} & 75.9{\fontsize{6pt}{6pt} \selectfont $\pm$0.2} & 39.3{\fontsize{6pt}{6pt} \selectfont $\pm$0.4} & 72.1 \\
BEATs$_\text{iter3+}$ \cite{chen2022beats} & (\checkmark) & 95.5{\fontsize{6pt}{6pt} \selectfont $\pm$0.3} & 87.6{\fontsize{6pt}{6pt} \selectfont $\pm$0.3} & 86.7{\fontsize{6pt}{6pt} \selectfont $\pm$0.1} & 37.0{\fontsize{6pt}{6pt} \selectfont $\pm$0.2} & 92.5{\fontsize{6pt}{6pt} \selectfont $\pm$0.1} & 67.6{\fontsize{6pt}{6pt} \selectfont $\pm$1.5} & 84.6{\fontsize{6pt}{6pt} \selectfont $\pm$0.5} & 73.1{\fontsize{6pt}{6pt} \selectfont $\pm$0.4} & 35.7{\fontsize{6pt}{6pt} \selectfont $\pm$0.3} & 73.4 \\
CED \cite{dinkel2023ced} &  & 97.3{\fontsize{6pt}{6pt} \selectfont $\pm$0.5} & 87.8{\fontsize{6pt}{6pt} \selectfont $\pm$0.2} & 89.0{\fontsize{6pt}{6pt} \selectfont $\pm$0.3} & 35.2{\fontsize{6pt}{6pt} \selectfont $\pm$0.2} & 94.8{\fontsize{6pt}{6pt} \selectfont $\pm$0.1} & 66.1{\fontsize{6pt}{6pt} \selectfont $\pm$1.3} & 42.3{\fontsize{6pt}{6pt} \selectfont $\pm$15.4} & 75.6{\fontsize{6pt}{6pt} \selectfont $\pm$0.5} & 38.9{\fontsize{6pt}{6pt} \selectfont $\pm$0.6} & 69.7 \\
AST\cite{gong2021ast} & \checkmark & 94.6{\fontsize{6pt}{6pt}\selectfont$\pm$0.7} & 85.4{\fontsize{6pt}{6pt}\selectfont$\pm$0.6} & 72.6{\fontsize{6pt}{6pt}\selectfont$\pm$0.0} & 16.5{\fontsize{6pt}{6pt}\selectfont$\pm$0.3} & 81.2{\fontsize{6pt}{6pt}\selectfont$\pm$0.3} & 58.4{\fontsize{6pt}{6pt}\selectfont$\pm$3.1} & {85.1{\fontsize{6pt}{6pt}\selectfont$\pm$0.5}}& 72.9{\fontsize{6pt}{6pt}\selectfont$\pm$0.9} & 25.7{\fontsize{6pt}{6pt}\selectfont$\pm$0.1} & 65.8 \\
HTS-AT \cite{Chen2022HTS-AT} & \checkmark & 95.7{\fontsize{6pt}{6pt} \selectfont $\pm$0.7} & 83.8{\fontsize{6pt}{6pt} \selectfont $\pm$0.1} & 82.1{\fontsize{6pt}{6pt} \selectfont $\pm$0.3} & 18.1{\fontsize{6pt}{6pt} \selectfont $\pm$0.4} & 82.3{\fontsize{6pt}{6pt} \selectfont $\pm$0.3} & 56.2{\fontsize{6pt}{6pt} \selectfont $\pm$0.6} & {85.1{\fontsize{6pt}{6pt} \selectfont $\pm$0.5}}& 73.3{\fontsize{6pt}{6pt} \selectfont $\pm$0.8} & 26.3{\fontsize{6pt}{6pt} \selectfont $\pm$0.5} & 67.0 \\
PANNs {\scriptsize CNN14}\cite{kong2020panns} & \checkmark & 90.7{\fontsize{6pt}{6pt}\selectfont$\pm$0.8} & 81.9{\fontsize{6pt}{6pt}\selectfont$\pm$0.1} & 51.8{\fontsize{6pt}{6pt}\selectfont$\pm$0.3} & 7.8{\fontsize{6pt}{6pt}\selectfont$\pm$0.1} & 75.1{\fontsize{6pt}{6pt}\selectfont$\pm$0.2} & 51.8{\fontsize{6pt}{6pt}\selectfont$\pm$1.3} & 78.7{\fontsize{6pt}{6pt}\selectfont$\pm$3.0} & 65.9{\fontsize{6pt}{6pt}\selectfont$\pm$0.2} & 10.9{\fontsize{6pt}{6pt}\selectfont$\pm$0.3} & 57.2 \\

\multicolumn{10}{l}{\textit{(Previous studies: Audio-language models)}}  \\
\addlinespace[0.05cm]
LAION-CLAP \cite{LAION-CLAP} & \checkmark & 97.3{\fontsize{6pt}{6pt} \selectfont $\pm$0.5} & 86.9{\fontsize{6pt}{6pt} \selectfont $\pm$0.5} & 75.9{\fontsize{6pt}{6pt} \selectfont $\pm$0.5} & 13.4{\fontsize{6pt}{6pt} \selectfont $\pm$0.4} & 80.3{\fontsize{6pt}{6pt} \selectfont $\pm$0.2} & 54.6{\fontsize{6pt}{6pt} \selectfont $\pm$1.0} & 84.3{\fontsize{6pt}{6pt} \selectfont $\pm$2.6} & 72.2{\fontsize{6pt}{6pt} \selectfont $\pm$1.1} & 14.8{\fontsize{6pt}{6pt} \selectfont $\pm$0.5} & 64.4 \\
CLAP$_{2022}$ \cite{CLAP2022} & \checkmark & 93.8{\fontsize{6pt}{6pt} \selectfont $\pm$0.1} & 84.2{\fontsize{6pt}{6pt} \selectfont $\pm$0.7} & 59.0{\fontsize{6pt}{6pt} \selectfont $\pm$1.1} & 8.9{\fontsize{6pt}{6pt} \selectfont $\pm$0.6} & 75.8{\fontsize{6pt}{6pt} \selectfont $\pm$1.3} & 54.4{\fontsize{6pt}{6pt} \selectfont $\pm$0.8} & 79.3 $\Lsh$ & 68.2{\fontsize{6pt}{6pt} \selectfont $\pm$0.6} & 8.4{\fontsize{6pt}{6pt} \selectfont $\pm$0.7} & 59.1 \\
CLAP$_{2023}$ \cite{CLAP2023} & \checkmark & {97.7{\fontsize{6pt}{6pt} \selectfont $\pm$0.5}}& 88.4{\fontsize{6pt}{6pt} \selectfont $\pm$0.1} & 86.2{\fontsize{6pt}{6pt} \selectfont $\pm$0.8} & 21.1{\fontsize{6pt}{6pt} \selectfont $\pm$0.3} & 89.6{\fontsize{6pt}{6pt} \selectfont $\pm$0.8} & 62.5{\fontsize{6pt}{6pt} \selectfont $\pm$1.8} & 82.3{\fontsize{6pt}{6pt} \selectfont $\pm$0.5} &\textbf{80.5{\fontsize{6pt}{6pt} \selectfont $\pm$0.1}}& 27.2{\fontsize{6pt}{6pt} \selectfont $\pm$0.5} & 70.6 \\
Pengi \cite{deshmukh2023Pengi} & \checkmark & 89.15 $\Lsh$ & - & - & - & - & 50.57 $\Lsh$ & 80.0 $\Lsh$ & - & - & - \\
WavCaps \cite{Mei2023WavCaps} & \checkmark & 97.2{\fontsize{6pt}{6pt} \selectfont $\pm$0.3} & 63.6{\fontsize{6pt}{6pt} \selectfont $\pm$0.6} & 73.3{\fontsize{6pt}{6pt} \selectfont $\pm$1.7} & 16.9{\fontsize{6pt}{6pt} \selectfont $\pm$0.2} & 80.0{\fontsize{6pt}{6pt} \selectfont $\pm$1.0} & 58.6{\fontsize{6pt}{6pt} \selectfont $\pm$0.7} & 80.2{\fontsize{6pt}{6pt} \selectfont $\pm$1.3} & 74.4{\fontsize{6pt}{6pt} \selectfont $\pm$0.9} & 21.1{\fontsize{6pt}{6pt} \selectfont $\pm$0.2} & 62.8 \\
Cacophony \scriptsize{(stage 2)}\cite{Zhu24Cacophony} $\dagger$ & & 97.0 $\Lsh$ & - & 76.2 $\Lsh$ & - & - & 59.3 $\Lsh$ & 85.0 $\Lsh$ & - & - & - \\
CoLLAT\cite{silva2023collat} & \checkmark & 97 $\Lsh$ & 89 $\Lsh$ & - & - & - & - & - & - & - & - \\
\midrule
\multicolumn{6}{l}{\textit{(Baseline: M2D)}}  \\
\addlinespace[0.05cm]
M2D \cite{M2D2024TASLP} & & 91.3{\fontsize{6pt}{6pt} \selectfont $\pm$0.6} & 87.6{\fontsize{6pt}{6pt} \selectfont $\pm$0.2} & \textbf{96.0{\fontsize{6pt}{6pt} \selectfont $\pm$0.1}} & 73.4{\fontsize{6pt}{6pt} \selectfont $\pm$0.2} & 98.3{\fontsize{6pt}{6pt} \selectfont $\pm$0.0} & 73.0{\fontsize{6pt}{6pt} \selectfont $\pm$0.7} & 84.1{\fontsize{6pt}{6pt} \selectfont $\pm$2.7} & 75.7{\fontsize{6pt}{6pt} \selectfont $\pm$0.1} & 42.1{\fontsize{6pt}{6pt} \selectfont $\pm$0.2} & 80.2 \\
M2D$^\text{AS}$\cite{M2D2024TASLP} & \checkmark & 96.7{\fontsize{6pt}{6pt} \selectfont $\pm$0.5} & 89.3{\fontsize{6pt}{6pt} \selectfont $\pm$0.1} & 93.2{\fontsize{6pt}{6pt} \selectfont $\pm$0.2} & 52.7{\fontsize{6pt}{6pt} \selectfont $\pm$0.6} & 96.9{\fontsize{6pt}{6pt} \selectfont $\pm$0.0} & 69.5{\fontsize{6pt}{6pt} \selectfont $\pm$1.8} & \textbf{87.6{\fontsize{6pt}{6pt} \selectfont $\pm$0.9}} & 75.6{\fontsize{6pt}{6pt} \selectfont $\pm$0.6} & 40.5{\fontsize{6pt}{6pt} \selectfont $\pm$0.4} & 78.0 \\
\multicolumn{6}{l}{\textit{(Proposed: Audio-language model)}}  \\
\addlinespace[0.05cm]
M2D-CLAP$_\text{2025}$ & \checkmark &\textbf{97.9{\fontsize{6pt}{6pt} \selectfont $\pm$0.5}}&\textbf{89.7{\fontsize{6pt}{6pt} \selectfont $\pm$0.3}}& 94.8{\fontsize{6pt}{6pt} \selectfont $\pm$0.2} & 56.3{\fontsize{6pt}{6pt} \selectfont $\pm$0.5} & 97.1{\fontsize{6pt}{6pt} \selectfont $\pm$0.0} & 69.7{\fontsize{6pt}{6pt} \selectfont $\pm$0.3} & 86.3{\fontsize{6pt}{6pt} \selectfont $\pm$1.8} & 76.7{\fontsize{6pt}{6pt} \selectfont $\pm$0.4} & 41.1{\fontsize{6pt}{6pt} \selectfont $\pm$0.3} & 78.8 \\
\addlinespace[0.05cm]

\multicolumn{12}{l}{\textit{(Ablations: No fine-tuning to AudioSet. M2D-CLAP$^\text{ stage1}_\text{2024}$ is pre-trained using a non-LLM text encoder and audio data only from AudioSet.)}}  \\
\addlinespace[0.05cm]
M2D-CLAP$^\text{ stage1}_\text{2025}$ & & 95.2{\fontsize{6pt}{6pt} \selectfont $\pm$0.3} & 89.5{\fontsize{6pt}{6pt} \selectfont $\pm$0.2} & 95.6{\fontsize{6pt}{6pt} \selectfont $\pm$0.1} & 68.6{\fontsize{6pt}{6pt} \selectfont $\pm$0.5} & 98.1{\fontsize{6pt}{6pt} \selectfont $\pm$0.1} & \textbf{73.4{\fontsize{6pt}{6pt} \selectfont $\pm$0.7}} & 86.6{\fontsize{6pt}{6pt} \selectfont $\pm$1.6} & 76.5{\fontsize{6pt}{6pt} \selectfont $\pm$0.4} & 42.2{\fontsize{6pt}{6pt} \selectfont $\pm$0.3} & 80.6 \\
\addlinespace[0.05cm]

M2D-CLAP$^\text{ stage1}_\text{2024}$\cite{niizumi24M2D-CLAP} & & {96.3{\fontsize{6pt}{6pt} \selectfont $\pm$0.3}} & 88.8{\fontsize{6pt}{6pt} \selectfont $\pm$0.6} & {95.8{\fontsize{6pt}{6pt} \selectfont $\pm$0.3}} & {70.3{\fontsize{6pt}{6pt} \selectfont $\pm$0.4}} & {98.3{\fontsize{6pt}{6pt} \selectfont $\pm$0.1}} & {\textbf{73.4{\fontsize{6pt}{6pt} \selectfont $\pm$0.2}}} & 84.1{\fontsize{6pt}{6pt} \selectfont $\pm$1.5} & {78.0{\fontsize{6pt}{6pt} \selectfont $\pm$0.5}} & {\textbf{42.4{\fontsize{6pt}{6pt} \selectfont $\pm$0.6}}} & {\textbf{80.8}}\\

\bottomrule
\addlinespace[0.08cm]
\multicolumn{10}{l}{$\Lsh$ Results quoted from corresponding papers when they are better than ours or unavailable in our test.}\\
\multicolumn{10}{l}{$^\dagger$ Results using the HEAR benchmark\cite{turian2022hear} in the settings similar to those for the linear evaluation.}\\
\end{tabular}
}
\vspace{-10pt}
\end{table*}

\subsubsection{Evaluation tasks}
Table \ref{tab:list-ds} lists the downstream tasks, which include environmental sound tasks (AudioSet\cite{gemmeke2017audioset}, ESC-50\cite{piczak2015esc50}, UrbanSound8K\cite{salamon2014urbansound} (US8K), and FSD50K), speech tasks (Speech Commands V2\cite{speechcommandsv2} (SPCV2), VoxCeleb1\cite{voxceleb} (VC1), VoxForge\cite{voxforge} (VF), CREMA-D\cite{cao2014cremad} (CRM-D)), and music tasks (GTZAN\cite{gt2013gtzan}, NSynth\cite{nsynth2017}, and Pitch Audio Dataset (Surge synthesizer)\cite{turian2021torchsynth} (Surge)).
AudioSet has two settings: AS2M with all 2M training samples and AS20K with 21K samples from the balanced train segments, and Surge is a non-semantic pitch audio classification of 88 MIDI notes.

All tasks are classification setups, and results are accuracies or mean average precision (mAP). When data contained variable-length audio, we randomly cropped or zero-padded the average audio length for each sample to create a batch of length-aligned audio fed into the model as in \cite{niizumi2023byol-a}\cite{M2D2024TASLP}.

We also used AudioCaps\cite{kim2019audiocaps} and Clotho\cite{icassp20Clotho} for audio-text retrieval and audio captioning. In addition, we used various music tasks supported by the MARBLE benchmark\cite{MARBLE}.

\begin{table*}[htb!]
%\vspace{-15pt}
\caption{Frozen audio feature evaluation results on music tasks under MARBLE benchmark.}
\label{tab:results-marble}
\centering
%\vspace{-5pt}
\resizebox{0.95\textwidth}{!}{%
\begin{tabular}{lccccccccccccccccc} \toprule
  & \multicolumn{2}{c}{MTT} & GTZAN & \multicolumn{2}{c}{EMO} & \multicolumn{2}{c}{NSynth} & \multicolumn{2}{c}{VocalSet} & \multicolumn{8}{c}{MTG} \\
  & \multicolumn{2}{c}{Tagging} & Genre & \multicolumn{2}{c}{Emotion} & Inst. & Pitch & Tech & Singer & \multicolumn{2}{c}{Instrument} & \multicolumn{2}{c}{MoodTheme} & \multicolumn{2}{c}{Genre} & \multicolumn{2}{c}{Top50} \\
 \cmidrule(lr){2-3} \cmidrule(lr){4-4} \cmidrule(lr){5-6} \cmidrule(lr){7-8} \cmidrule(lr){9-10} \cmidrule(lr){11-18}
Model & ROC & AP & Acc & R2$^V$ & R2$^A$ & Acc & Acc & Acc & Acc & ROC & AP & ROC & AP & ROC & AP & ROC & AP\\
\midrule
\multicolumn{10}{l}{\textit{(Previous studies: General audio and audio-language models)}}  \\
\addlinespace[0.05cm]
ATST-Frame\cite{Li2023ATST-TALSP} & 91.2 & 39.7 & 79.0 & 60.3 & 76.7 & 77.0 & 84.2 & 76.8 & 88.7 & 78.5 & 21.0 & 77.0 & 15.1 & 87.0 & 19.5 & 83.0 & 29.7 \\
BEATs$_\textbf{iter3+}$\cite{chen2022beats} & 91.6 & 40.7 & 84.6 & 55.1 & 70.9 & 78.1 & 84.2 & 74.8 & 84.8 & \multicolumn{8}{c}{\textit{N/A (unable to handle long audio)}} \\
CLAP$_{2023}$\cite{CLAP2023} & 90.0 & 38.6 & 79.0 & 47.6 & 68.1 &\textbf{81.0}& 68.1 & 70.7 & 73.7 & 76.2 & 18.5 & 75.4 & 12.9 & 83.9 & 15.6 & 80.8 & 26.2 \\

\multicolumn{10}{l}{\textit{(Previous studies: SOTA music audio models)}}  \\
\addlinespace[0.05cm]
Jukebox-5B\cite{dhariwal2020jukebox}  & 91.4 & 40.6 & 77.9 & 57.0 & 73.0 & 70.4 & 91.6 & 76.7 & 82.6 & 78.5 & 22.0 & 77.6 & 15.3 & 88.0 & 20.5 & 83.7 & 30.6 \\
MULE\cite{ISMIR22MULE}     & 91.2 & 40.1 & 75.5 & 60.7 & 73.1 & 74.6 & 88.5 & 75.5 & 87.5 & 76.6 & 19.2 & 78.0 & 15.4 & 88.0 & 20.4 & 83.7 & 30.6 \\
MAP-MERT-v1-95M\cite{Li2024MERT} & 91.0 & 39.3 & 74.8 & 55.5 & 76.3 & 70.7 & 92.6 & 74.2 & 83.7 & 77.5 & 19.4 & 76.4 & 13.4 & 87.1 & 18.8 & 83.0 & 29.0 \\
MAP-MERT-v1-330M\cite{Li2024MERT} & 91.1 & 39.5 & 77.6 & 59.0 & 75.8 & 72.6 & \textbf{94.4} & 76.9 & 87.1 & 78.1 & 19.8 & 76.5 & 14.0 & 86.7 & 18.6 & 83.4 & 29.9 \\

\midrule
\multicolumn{10}{l}{\textit{(Baseline: M2D)}}  \\
\addlinespace[0.05cm]
M2D & 91.4 & 40.5 & 81.5 & 59.4 & 76.1 & 78.2 & 89.3 & 76.1 & 92.0 & 79.2 & 21.1 & 77.6 & 15.7 & 87.6 & 20.1 & 83.4 & 29.8 \\
M2D$^\text{AS}$ & 91.5 & 40.9 &\textbf{86.1}& 59.0 & 73.0 & 78.7 & 88.2 & 76.8 & 91.3 &\textbf{79.7}&\textbf{23.3}& 77.8 & 15.6 & 87.5 & 20.3 & 83.4 & 30.6 \\

\multicolumn{10}{l}{\textit{(Proposed: General audio-language model)}}  \\
\addlinespace[0.05cm]
M2D-CLAP$_\text{2025}$  &\textbf{91.8}&\textbf{41.6}& {85.9} & 58.2 & 76.2 & 79.7 & 89.5 &\textbf{78.9}&\textbf{92.7}& \textbf{79.7} & {22.2} &\textbf{78.8}& 15.8 &\textbf{88.3}&\textbf{20.9}&\textbf{84.0}&\textbf{30.9} \\
\addlinespace[0.05cm]
 
\multicolumn{18}{l}{\textit{(Ablations: No fine-tuning to AudioSet. M2D-CLAP$^\text{ stage1}_\text{2024}$ is pre-trained using a non-LLM text encoder and audio data only from AudioSet.)}}  \\

M2D-CLAP$^\text{ stage1}_\text{2025}$ & 91.5 & 40.8 & 82.6 & 59.3 & 76.7 & 77.4 & 89.1 & 77.4 & 91.8 & 79.3 & 21.7 & 78.0 & 15.9 & 87.9 & 20.8 & 83.7 & 30.3 \\
\addlinespace[0.05cm]

M2D-CLAP$^\text{ stage1}_\text{2024}$\cite{niizumi24M2D-CLAP} & {91.6} & {41.0} & {84.3} &{\textbf{61.9}}&{\textbf{77.4}}& {80.6} & 88.8 & {78.1} & {92.3} & 79.0 & 21.6 & {78.2} &{\textbf{16.0}}& {88.0} & 20.8 & {83.8} & {30.4} \\

\bottomrule
\end{tabular}
}
\vspace{-10pt}
\end{table*}

\subsection{Evaluation: Audio Features} \label{sec:exp-audio-feature}
We evaluated the effectiveness of the pre-trained audio features under the frozen feature and fine-tuning settings on general tasks and the frozen feature setting on music tasks.
We followed \cite{niizumi2023byol-a}\cite{M2D2024TASLP} for the frozen feature (linear evaluation) and fine-tuning experimental setup using a unified evaluation platform, EVAR\footnote{\url{https://github.com/nttcslab/eval-audio-repr}}, for general tasks. We also evaluated features for music tasks using the MARBLE benchmark\cite{MARBLE}.

\subsubsection{Frozen audio features on general audio tasks} \label{sec:exp-audio-LE}

\paragraph{Experimental setup}
We followed the standard linear evaluation procedure\cite{niizumi2023byol-a}\cite{M2D2024TASLP}. Specifically, the evaluation pipeline first converts downstream task samples into features using a \textit{frozen} pre-trained model, trains a linear layer with task labels, and then conducts tests for the trained linear layer. 
We used the validation set for early stopping with patience of 20 epochs and trained the linear layer for up to 200 epochs using the Adam optimizer and a learning rate of 0.00003.

\paragraph{Results: Frozen audio features on general audio tasks}
The linear evaluation results in Table \ref{tab:results-le} show that M2D and M2D-CLAP variants perform well across the tasks, whereas supervised learning models (AST, HTS-AT, and PANNs) and CLAP models (e.g., LAION-CLAP) perform poorly on speech tasks and Surge. For instance, their scores range from 7.8 to 18.1 on VC1 for supervised models and 8.4 to 27.2 on Surge for CLAP models. In contrast, M2D and M2D-CLAP variants achieve significantly higher scores, ranging from 52.7 to 73.4 on VC1 and from 40.5 to 42.4 on Surge.
Note that the CLAP models employ backbone audio encoders trained with supervised learning, which limits the generalizability of their audio features to tasks outside the scope of their training labels and captions. For example, a label such as \textit{"Male speech, man speaking"} does not capture speaker identity, as required in VC1 (speaker identification task).

BEATs and CED, pre-trained with distillation of the previous training cycle models, show a similar trend to supervised learning models. CED shows 35.2 on VC1, and BEATs$_\text{iter3+}$ shows 35.7 on Surge, for example. These models have distilled the pre-trained models on AudioSet, making them focused more on AudioSet data distribution. This specialization may hinder their generalization to tasks involving different data distributions.

In contrast, M2D, M2D-CLAP, and an SSL model, ATST, perform well in all tasks, and M2D-CLAP$_\text{2025}$ achieves SOTA performance on ESC-50 at 97.9 and US8K at 89.7.
However, fine-tuning on AudioSet may have reduced the sensitivity of the representations to fine-grained speaker differences, causing a performance drop in M2D-CLAP$_\text{2025}$ on VC1 (56.3 compared to 73.4 with M2D), although 56.3 still remains a competitive performance.
The first-stage model, M2D-CLAP$^\text{ stage1}_\text{2025}$, also shows slightly lower performance on speech tasks and reduces the overall average performance compared to M2D, suggesting that the use of AudioSet captions or label supervision may be the cause.

In addition, M2D-CLAP$^\text{ stage1}_\text{2024}$, which is pre-trained only on AudioSet using GTE-Base\cite{li2023GTE} (non-LLM-based sentence embedding model) as a text encoder, performs better than M2D-CLAP$^\text{ stage1}_\text{2025}$ in most tasks, indicating that the pre-training setting, especially relying solely on AudioSet is beneficial for performing well with frozen audio features. However, as shown in Section \ref{sec:eval-clap-feature}, its CLAP features severely underperform M2D-CLAP$^\text{ stage1}_\text{2025}$.

In summary, M2D-CLAP maintains strong performance as a frozen audio feature extractor on general audio tasks, albeit with compromised performance on speech tasks.

\subsubsection{Frozen audio features on music tasks} \label{sec:exp-audio-music}
To further assess the frozen feature, we evaluated models on the music audio tasks with various types of audio signals.

\paragraph{Experimental setup}
We used the MARBLE benchmark\cite{MARBLE}, which covers diverse music audio tasks for music tracks, instrument sounds, and vocals. We selected the tasks for which we could reproduce the results with the released code\footnote{\url{https://github.com/a43992899/MARBLE-Benchmark}}. The tasks include music tagging and genre classification on MagnaTagATune (MTT)\cite{MagnaTagATune}, GTZAN\cite{gt2013gtzan}, and MTG-Jamendo (MTG)\cite{MTG-Jamendo}, emotion detection on Emomusic\cite{Emomusic} and MTG, pitch and instrument (inst) classification on NSynth\cite{nsynth2017} and MTG, and vocal technique detection (tech) and singer identification on VocalSet\cite{VocalSet}.
The evaluation metrics are the area under the ROC curve (ROC) and the average precision (AP), except for Emomusic, which uses the determination coefficient between the model’s regression results and human annotations of valence (R2$^V$) and arousal (R2$^A$)\cite{Emomusic}.
The results are the averages of five tests with different random seeds.

\paragraph{Results: Frozen audio features on music tasks}
The results of the music task benchmark in Table \ref{tab:results-marble} show that general audio models perform comparably to the SOTA music audio models. They even outperform music models on GTZAN and NSynth (Inst) while underperforming them in pitch classification NSynth (Pitch).
While music models are pre-trained on large-scale music databases, general models are also pre-trained using large-scale music data because more than half ($>$1M) of the AudioSet samples are tagged as "music"\cite{gemmeke2017audioset}. 
In addition, most of the music and general models share a similar SSL-based training scheme. These facts would explain the comparable performance of general models on the music-specific benchmark.
However, CLAP$_{2023}$, which uses a supervised learning-based audio backbone, underperforms other general and music models (except NSynth (Inst)\footnote{The CLAP$_{2023}$ training data contains NSynth.}), suggesting that SSL on AudioSet is important for generalizing to music tasks, as demonstrated by other SSL-based general audio models.

The models pre-trained using both SSL and supervised learning of AudioSet (M2D-CLAP$_\text{2025}$ and BEATs$_\text{iter3+}$) show stronger performance, and M2D-CLAP$_\text{2025}$ demonstrates new SOTA results on MTT, VocalSet, and MTG.
However, these models underperform music audio models on the NSynth (Pitch) task, indicating that music models have higher resolution in capturing frequency-related information than general models. Notably, M2D-CLAP achieves performance on par with MULE, a music model, on this task.
In addition, as in the evaluation on general tasks in Table \ref{tab:results-le}, M2D-CLAP$^\text{ stage1}_\text{2024}$ performs better than M2D-CLAP$^\text{ stage1}_\text{2025}$ in most tasks, also indicating that pre-training solely on AudioSet is beneficial for performing well with frozen features.

To summarize, the experiment confirmed that M2D-CLAP performs well and shows SOTA performance in music tasks.

\subsubsection{Fine-tuning} \label{sec:exp-audio-FT}
For fine-tuning, we focus primarily on evaluating M2D-CLAP$^\text{ stage1}_\text{2025}$.  We discuss the final M2D-CLAP$_\text{2025}$ as an ablation model because it has already been fine-tuned to AudioSet once in the additional stage 1.1.

\paragraph{Experimental setup}
We followed M2D\cite{M2D2024TASLP}, ATST\cite{Li2023ATST-TALSP}, and Audio-MAE\cite{huang2022amae} for the fine-tuning procedure.
Specifically, a linear classifier was added on top of the pre-trained model to train the entire network.
We used SpecAugment\cite{specaugment}, Mixup\cite{zhang2018mixup,niizumi2023byol-a}, and Random Resize Crop (RRC)\cite{niizumi2023byol-a} for data augmentation, along with Structured Patchout\cite{Koutini2022passt} that randomly masks patches during training. We used SGD, LARS, AdamW optimizers, and cosine annealing\cite{loshchilov2016sgdr} learning rate scheduling.
We used the hyperparameters for each task as listed in Table \ref{tab:exp-general:ft-parameters}.
For Structured Patchout, we used 768-d features calculated by averaging all ViT outputs because Eq. \eqref{eq:msm-mae} is not applicable to the masked patches.
We used the tasks commonly used in previous studies, namely ESC-50, SPCV2, and VC1, the same as in the linear evaluation, plus AS2M and AS20K.

Unlike the linear evaluation covering most models, we limited fine-tuning evaluation to our models, as reproducing previous best results often involves non-trivial tuning.

\begin{table}[tb!]
%\vspace{-5pt}
\caption{Fine-tuning evaluation parameters.}
\label{tab:exp-general:ft-parameters}
\vspace{-5pt}
\centering
\resizebox{1.0\columnwidth}{!}{%
\begin{tabular}{llllll}
\toprule
Parameter & AS2M & AS20K & ESC-50 & SPCV2 & VC1 \\
\midrule
Learning rate & 2.0 & 0.5 & 0.5 & 0.5 & 0.0005 \\
Batch size & 64 & 64 & 128 & 128 & 64 \\
Optimizer & LARS & SGD & SGD & SGD & AdamW \\
Mixup\cite{zhang2018mixup,niizumi2023byol-a} ratio & 0.5 & 0.3 & 0.0 & 0.3 & 0.0 \\
Random resize crop & - & \checkmark & \checkmark & \checkmark & - \\
SpecAugment$^\dagger$\cite{specaugment} & 30/192 & 30/192 & 15/48 & 30/48 & 30/48 \\
Training epochs (total) & 70 & 200 & 200 & 200 & 50 \\
Training epochs (warm-up) & 15 & 5 & 5 & 5 & 5 \\
Structured Patchout\cite{Koutini2022passt} ratio & 0.5 & 0.5 & 0.5 & 0.5 & 0.0 \\
Adjust positional encoding & \checkmark & \checkmark & - & - & \checkmark \\
Freeze embedding layer\cite{kumar2022finetune} & - & - & \checkmark & - & - \\
\bottomrule
\addlinespace[0.05cm]
\multicolumn{6}{l}{$^{\dagger}$ Frequency/time masking parameters.}\\
\end{tabular}
}
\vspace{-5pt}
\end{table}

\begin{table}[tb!]
%\vspace{-5pt}
\caption{Fine-tuning evaluation results with 95\% CI.}
\label{tab:exp-general:audio-task-ft}
\vspace{-5pt}
\centering
\resizebox{1.0\columnwidth}{!}{%
\begin{tabular}{llllll}
\toprule
 & AS2M & AS20K &     ESC-50 &  SPCV2 &       VC1\\
\vspace{-1pt} Model  & mAP & mAP &  acc(\%) &  acc(\%) &    acc(\%)\\
\midrule
\multicolumn{6}{l}{\textit{(Previous studies: Audio models)}}  \\
BEATs$_\text{iter3}$ \cite{chen2022beats} & 48.0 & 38.3 & 95.6 & 98.3 & - \\
BEATs$_\text{iter3+}$ \cite{chen2022beats} & 48.6 & 41.8 & 98.1& 98.1 & - \\
ATST-Clip \cite{Li2023ATST-TALSP} & 45.2 & 37.9 & - & 98.0 & 95.5 \\
ATST-Frame \cite{Li2023ATST-TALSP} & 48.0 & 39.0 & - & 98.1 & \textbf{97.3}\\
HTS-AT \cite{Chen2022HTS-AT} & 47.1 & - & 97.0 & 98.0 & - \\
DTF-AT \scriptsize{Full TF}\cite{Ahmed2024DTF-AT} & 48.6 & - & 97.5 & 97.68 & - \\
\multicolumn{6}{l}{\textit{(Previous studies: Audio-language models)}}  \\
\addlinespace[0.05cm]
AudioCLIP \cite{AudioCLIP} & - & - & 97.15 & - & - \\
Wav2CLIP \cite{Wav2CLIP} & - & - & 85.95 & - & - \\
CLAP$_{2022}$ \cite{CLAP2022} & - & - & 96.7 & 96.8 & - \\
\midrule
\multicolumn{6}{l}{\textit{(Baseline: M2D)}}  \\
M2D & {47.9 {\fontsize{6pt}{6pt}\selectfont $\pm$ 0.0}}&{38.6 {\fontsize{6pt}{6pt}\selectfont $\pm$ 0.1}}&{96.0 {\fontsize{6pt}{6pt}\selectfont $\pm$ 0.2}}& \textbf{98.4 {\fontsize{6pt}{6pt}\selectfont $\pm$ 0.1}} & 96.3 {\fontsize{6pt}{6pt}\selectfont $\pm$ 0.2} \\

\multicolumn{6}{l}{\textit{(Proposed: Audio-Language model)}}  \\
M2D-CLAP$^\text{ stage1}_\text{2025}$ &\textbf{49.0{\fontsize{6pt}{6pt} \selectfont $\pm$0.1}}& 42.1{\fontsize{6pt}{6pt} \selectfont $\pm$0.0} & 97.8{\fontsize{6pt}{6pt} \selectfont $\pm$0.1} & \textbf{98.4{\fontsize{6pt}{6pt} \selectfont $\pm$0.0}} & 95.5{\fontsize{6pt}{6pt} \selectfont $\pm$0.3} \\
\addlinespace[0.05cm]

\multicolumn{6}{l}{\textit{(Ablations: M2D-CLAP$_\text{2025}$ have been fine-tuned to AudioSet in stage 1.1.)}}  \\
\addlinespace[0.05cm]
M2D-CLAP$_\text{2025}$ & - & \textbf{46.7{\fontsize{6pt}{6pt} \selectfont $\pm$0.2}} & \textbf{98.5{\fontsize{6pt}{6pt} \selectfont $\pm$0.0}} & \textbf{98.4{\fontsize{6pt}{6pt} \selectfont $\pm$0.1}} & 94.0{\fontsize{6pt}{6pt} \selectfont $\pm$0.2} \\
\addlinespace[0.05cm]
M2D-CLAP$^\text{ stage1}_\text{2024}$\cite{niizumi24M2D-CLAP} & 48.5{\fontsize{6pt}{6pt} \selectfont $\pm$0.1} & 41.8{\fontsize{6pt}{6pt} \selectfont $\pm$0.2} & 97.4{\fontsize{6pt}{6pt} \selectfont $\pm$0.2} & 98.3{\fontsize{6pt}{6pt} \selectfont $\pm$0.1} & 95.5{\fontsize{6pt}{6pt} \selectfont $\pm$0.2} \\

\midrule

\multicolumn{6}{l}{\textit{(Reference: Models using ensemble of audio models or distillation from large models)}}  \\
CED \cite{dinkel2023ced}  &{\color{EMgray}50.0}&{\color{EMgray}44.0}& {\color{EMgray}96.65} & - & - \\
ATST-C2F \cite{Li2023ATST-TALSP}  & {\color{EMgray}49.7} & {\color{EMgray}40.5} & - & {\color{EMgray}98.4} &{\color{EMgray}97.5}\\
SupMAM-CLAP \cite{xin23d_mamclap}  & {\color{EMgray}48.5} & {\color{EMgray}38.6} & {\color{EMgray}97.6} & {\color{EMgray}98.7} & - \\
DyMN-L\cite{Schmid2024DyMN} & {\color{EMgray}49.0}& - & {\color{EMgray}97.4} & - & - \\
\bottomrule
\end{tabular}
}
\vspace{-5pt}
\end{table}

\paragraph{Results: Fine-tuning}
Table \ref{tab:exp-general:audio-task-ft} shows that M2D-CLAP$^\text{ stage1}_\text{2025}$ improves upon the baseline M2D, achieving SOTA performance on AS2M (49.0) and strong results on AS20K (42.1) and ESC-50 (97.8), outperforming the previous top model, BEATs$\text{iter3+}$, except on ESC-50.
However, VC1 performance degrades to 95.5, lower than the original M2D's performance (96.3), suggesting that learning from the CLAP supervision sacrifices fine-grained feature representation.

Compared to ablation M2D-CLAP$^\text{ stage1}_\text{2024}$, M2D-CLAP$^\text{ stage1}_\text{2025}$ shows superior performance on all tasks, suggesting that the effect of the semantic supervision of LLM-based sentence embeddings is effective in fine-tuning,  but does not contribute as a frozen feature, as confirmed in Section \ref{sec:exp-audio-LE}.

The final M2D-CLAP$_\text{2025}$, which we handle as ablation, shows improved performance of 46.7 for AS20K and 98.5 for ESC-50, while 94.0 for VC1, a -1.5 pp degradation compared to M2D-CLAP$^\text{ stage1}_\text{2025}$. The degradation of VC1 reaffirms the trade-off caused by fine-tuning under AudioSet's supervision.

In summary, M2D-CLAP provides top-performing audio features for fine-tuning. Notably, this is achieved by a single pre-training process, unlike other SOTA models that require multiple stages or models.
The first stage model M2D-CLAP$^\text{ stage1}_\text{2025}$ also shows consistent performance gains across other experiments in Section \ref{sec:exp-audio-feature}.

\subsection{Evaluation: CLAP Features} \label{sec:eval-clap-feature}
We evaluated the effectiveness of the CLAP features in zero-shot classification, audio-text retrieval, and audio captioning experiments.

\subsubsection{Zero-shot Classification} \label{sec:exp-clap-ZS}
We conducted experiments to evaluate the zero-shot classification performance of CLAP features. In this setting, the model, without any task-specific training, predicts the class label of an audio clip by selecting the one whose CLAP text feature has the highest similarity to the CLAP audio feature.

\paragraph{Experimental setup}
We chose the tasks found in previous papers, which include ESC-50 (ESC), US8K, CREMA-D (CRD), GTZAN (GTZ), NSynth (NS), AudioSet (AS), and FSD50K (FSD). We used AudioSet and FSD50K as references because M2D-CLAP was pre-trained on AudioSet and partially on FSD50K in WavCaps.
We conducted experiments using a standard procedure to infer a class label of audio with the closest cosine distance between the features of each test sample and the label's caption, and we obtained accuracies from the inference results.
Table \ref{tab:exp:zs-rules} summarizes the rules for converting task labels to the captions.
We used the evaluation platform EVAR, as in Section \ref{sec:exp-audio-feature}.

\paragraph{Results: Zero-shot classification}
Table \ref{tab:results-zs} shows the experimental results. We compare results based on a strict zero-shot classification setting as in \cite{Mei2023WavCaps}; we only include experimental results that do not use data from downstream tasks during training in our comparisons, and all other results are grayed out as references.

Results show that M2D-CLAP variants perform well on all tasks, and M2D-CLAP$_\text{2025}$ achieves top-level performance in all tasks.
Among the ablation variants, M2D-CLAP$^\text{ stage1}_\text{2025}$ outperforms previous models with a SOTA result of 79.31 for GTZAN, while the second-stage pre-training degrades performance, such as 72.41 with M2D-CLAP$^\text{ stage2}_\text{2025}$.

M2D-CLAP$^\text{ stage2.1}_\text{2025}$ brings the ESC-50 and US8K results to the top level, indicating that CLAP training with the limited combination of WavCaps, AudioCaps, and Clotho is advantageous for these tasks compared to the models M2D-CLAP$^\text{ stage1 \& 2}_\text{2025}$ that use large-scale datasets. Furthermore, M2D-CLAP$^\text{ stage1}_\text{2025}$ significantly improves performance on these tasks from M2D-CLAP$^\text{ stage1}_\text{2024}$\cite{niizumi24M2D-CLAP}, indicating the benefit of using an LLM-based text encoder as well as incorporating additional data from VGGSound and WavCaps.

In summary, we validated that M2D-CLAP achieves top-level performance in the zero-shot classification setting.

\begin{table}[tb!]
%\vspace{-5pt}
\caption{Zero-shot caption conversion rules.}
\label{tab:exp:zs-rules}
\vspace{-5pt}
\centering
\resizebox{0.98\columnwidth}{!}{%
\begin{tabular}{ll}
\toprule
Task & Rule \\
\midrule
AS \& FSD & \textit{"$\langle$labels$\rangle$ can be heard"}  \\
ESC50 \& US8K & \textit{"$\langle$label$\rangle$ can be heard"}  \\
CREMA-D & \textit{"(angry person talking $|$ someone talking in disgust}  \\
 & \hspace{0.1cm} \textit{$|$ someone talking with a sense of fear}  \\
 & \hspace{0.1cm} \textit{$|$ someone talking happily and joyfully}  \\
 & \hspace{0.1cm} \textit{$|$ someone talking calmly $|$ someone talking sadly) can be heard"}  \\
GTZAN & \textit{"$\langle$label$\rangle$ music can be heard"}  \\
NSynth & \textit{"the musical instrument sound of $\langle$label$\rangle$ can be heard"}  \\
\bottomrule
\addlinespace[0.05cm]
\end{tabular}
}
\vspace{-5pt}
\end{table}

\begin{table}[tb!]
%\vspace{-5pt}
\caption{Zero-shot classification results. Results using test task data during training are grayed out as a reference.}
\label{tab:results-zs}
\centering
\vspace{-5pt}
\resizebox{\columnwidth}{!}{%
\begin{tabular}{llllllll} \toprule
  & AS & FSD & ESC &  US8K & CRD & GTZ &  NS \\
\vspace{-1pt} Model & mAP & mAP & acc(\%) & acc(\%) & acc(\%) & acc(\%) & acc(\%)\\
\midrule
\multicolumn{6}{l}{\textit{(Previous studies)}}  \\
\addlinespace[0.05cm]

AudioCLIP \cite{AudioCLIP} & - & - & 69.40 $\Lsh$ & 68.78 $\Lsh$ & - & - & - \\
Wav2CLIP \cite{Wav2CLIP} & - & 3.02 $\Lsh$ & 41.4 $\Lsh$ & 40.44 $\Lsh$ & - & - & - \\
WavCaps \cite{Mei2023WavCaps} & {\color{EMgray}19.60} & {\color{EMgray}52.96} & {94.8} $\Lsh$ & 81.42 & 19.86 & 45.52 & 27.66 \\
LAION-CLAP \cite{LAION-CLAP} & - & {\color{EMgray} 45.85} & 91.0 $\Lsh$ & 77.0 $\Lsh$ & 23.08 & 47.24 & \textbf{35.28} \\
CLAP$_{2022}$ \cite{CLAP2022} & \textbf{5.8} $\Lsh$ & {\color{EMgray} 30.24 $\Lsh$} & 82.6 $\Lsh$ & 75.29 & 22.76 & 28.97 & 21.44 \\
CLAP$_{2023}$ \cite{CLAP2023} &  {\color{EMgray} 10.2 $\Lsh$} & {\color{EMgray} 48.5 $\Lsh$} & 93.90 $\Lsh$ & {82.3} $\Lsh$ &\textbf{30.0} $\Lsh$ & 58.4 $\Lsh$ &  {\color{EMgray} 58.08} \\
MGA-CLAP$^+$\cite{MGA-CLAP} & {\color{EMgray} 23.0 $\Lsh$} & {\color{EMgray} 54.5 $\Lsh$} & \textbf{94.9} $\Lsh$ & \textbf{83.7} $\Lsh$ & - & - & - \\
Cacophony\cite{Zhu24Cacophony} & - & - & 93.4 $\Lsh$ & 77.1 $\Lsh$ & - & - & -  \\
Pengi \cite{deshmukh2023Pengi} &  {\color{EMgray}16.35 $\Lsh$} & {\color{EMgray}46.76 $\Lsh$} & 91.95 $\Lsh$ & 71.85 $\Lsh$ & 18.46 $\Lsh$ & 35.25 $\Lsh$ & {\color{EMgray} 50.07 $\Lsh$}\\
LTU\cite{gong2023LTU}/-AS\cite{gong2023LTU-AS} &  {\color{EMgray}18.7 $\Lsh$} & {\color{EMgray}46.3} $\Lsh$ & 83.1 $\Lsh$ & - & - & 50.3 $\Lsh$ & - \\
CoLLAT\cite{silva2023collat} & {\color{EMgray}9 $\Lsh$} & 19 $\Lsh$ & 84 $\Lsh$ & 77 $\Lsh$ & - & - & - \\
GLAP\cite{Dinkel2025GLAP} & - & {\color{EMgray}40.9 $\Lsh$} & 88.8 $\Lsh$ & 78.9 $\Lsh$ & - & 69.6 $\Lsh$ & 31.3 $\Lsh$ \\
JMLA \cite{du2023JMLA} & - & - & - & - & - & 64.82 $\Lsh$ & - \\
\midrule
\multicolumn{6}{l}{\textit{(Proposed model)}}  \\
\addlinespace[0.05cm]
M2D-CLAP$_\text{2025}$ & {\color{EMgray}30.05} & {\color{EMgray}50.04} & 94.30 & 82.30 & 28.56 & 70.69 & 30.57 \\
\addlinespace[0.05cm]

\multicolumn{6}{l}{\textit{(Ablations)}}  \\
M2D-CLAP$^\text{ stage1}_\text{2025}$ & {\color{EMgray}23.15} & {\color{EMgray}43.43} & 90.07 & 76.52 & 17.41 & \textbf{79.31} & 31.76 \\
\addlinespace[0.05cm]
M2D-CLAP$^\text{ stage2}_\text{2025}$ & {\color{EMgray}28.27} & {\color{EMgray}46.76} & 93.67 & 81.92 & 23.50 & 72.41 & 24.57 \\
\addlinespace[0.05cm]
M2D-CLAP$^\text{ stage2.1}_\text{2025}$ &  {\color{EMgray}27.13} & {\color{EMgray}42.03} & 94.55 & {82.90} & 21.73 & 70.86 & 29.52 \\
\addlinespace[0.05cm]
M2D-CLAP$^\text{ stage1}_\text{2024}$\cite{niizumi24M2D-CLAP} & {\color{EMgray}27.24} & \textbf{40.82} & 75.45 & 72.40 & 17.73 & {75.17}& 23.39 \\

\midrule
\multicolumn{8}{l}{\textit{(Reference studies that employ additional techniques on top of existing CLAP models.)}}  \\
Proto-LC \cite{Kushwaha23ProtoLC} & - & {\color{EMgray}52} $\Lsh$ & {\color{EMgray}{96} $\Lsh$} & {\color{EMgray}73 $\Lsh$} & - & - & - \\
CD$_\text{Ontology}$ \cite{Olvera2024SoundDesc} & {\color{EMgray}19.98 $\Lsh$} & {\color{EMgray}50.74 $\Lsh$} & {\color{EMgray}{96.35} $\Lsh$} & {\color{EMgray}{90.17} $\Lsh$} & - & - & - \\
\bottomrule
\addlinespace[0.08cm]
\multicolumn{8}{l}{$\Lsh$ Results quoted from papers when they are better than our reproduction or test unavailable.}\\
\end{tabular}
}
\vspace{-10pt}
\end{table}

\begin{table*}[tb!]
%\vspace{-5pt}
\caption{Audio-to-Text/Text-to-Audio retrieval comparisons. Dataset are WC~(WavCaps), A~(AudioCaps), C~(Clotho), LA~(LAION-Audio-630K), LS~(Large-scale datasets such as AudioSet).}
\label{tab:exp-atr:results}
\vspace{-5pt}
\centering
\resizebox{1.0\textwidth}{!}{%
\begin{tabular}{lccllllllllllllllll}
\toprule
  & \multirow{4}{*}{\shortstack[c]{Final\\CLAP\\training\\dataset}} & \multirow{4}{*}{\shortstack[c]{Audio\\encoder\\AudioSet\\fine-tuning}} & \multicolumn{8}{c}{AudioCaps} & \multicolumn{8}{c}{Clotho} \\
 \cmidrule(lr){4-11} \cmidrule(lr){12-19}
  &  &  & \multicolumn{4}{c}{Text-to-Audio} & \multicolumn{4}{c}{Audio-to-Text} & \multicolumn{4}{c}{Text-to-Audio} & \multicolumn{4}{c}{Audio-to-Text} \\
 \cmidrule(lr){4-7} \cmidrule(lr){8-11}  \cmidrule(lr){12-15} \cmidrule(lr){16-19}
CLAP model &  &  & R@1 & R@5 & R@10 & \scriptsize{mAP@10} & R@1 & R@5 & R@10 & \scriptsize{mAP@10} & R@1 & R@5 & R@10 & \scriptsize{mAP@10} & R@1 & R@5 & R@10 & \scriptsize{mAP@10}\\
\midrule
\multicolumn{6}{l}{\textit{(Previous studies)}}  \\
HTSAT-BERT-ZS\cite{Mei2023WavCaps} & WC & \checkmark & 28.6 & 61.1 & 75.8 & - & 40.2 & 69.4 & 80.3 & - & 16.5 & 38.8 & 50.9 & - & 20.0 & 43.3 & 56.6 & - \\
HTSAT-BERT-PT\cite{Mei2023WavCaps} & WC+A+C & \checkmark & 39.7 & 74.5 & 86.1 & - & 51.7 & 82.3 & 90.6 & - & 19.5 & 45.2 & 58.2 & - & 23.4 & 50.9 & 63.4 & - \\
HTSAT-BERT-FT\cite{Mei2023WavCaps} & A or C & \checkmark &\textbf{42.2}& {76.5}& {87.1}& - & 54.6 & \textbf{85.2}& {92.4}& - & {19.7}& {45.7}& {59.4}& - & 26.9 & {52.6}& 64.90 & - \\
CNN14-BERT-FT\cite{Mei2023WavCaps} & A or C & \checkmark & 35.1 &70.0 &82.1 &-&45.7 &76.1 &87.7 &-& \textbf{21.5} & \textbf{47.9} & \textbf{61.9} &-& \textbf{27.1} & 52.7 & 66.3 & -\\
LAION-CLAP\cite{LAION-CLAP} & LS+LA+A+C & \checkmark & 35.1 & 71.9 & 83.7 & - & 44.2 & 80.8 & 90.3 & - & 16.9 & 41.6 & 54.4 & - & 24.4 & 49.3 & {65.7}& - \\
CLAP$_{2023}$\cite{CLAP2023} & LS+WC+A+C & \checkmark & 35.6 & - & - & {51.0}& 42.5 & - & - & {31.9}& 15.7 & - & - & 25.7 & 22.9 & - & - & {15.5}\\
MGA-CLAP$^*$\cite{MGA-CLAP} & WC+A+C & \checkmark & \textbf{42.2} & 74.9 & - & - & 53.7 & 84.3 & - & - & 20.8 & 45.0 & - & - & 26.5 & 54.1 & - & - \\
Cacophony\cite{Zhu24Cacophony} & LS+A+C  & & 41.0 & 75.3 & 86.4 &-& 55.3 & 83.6 & 92.4 &-& 20.2 & 45.9 & 58.8 &-& 26.5 & \textbf{54.1} & 67.3 &- \\
FLAP$_\text{(+LLM-aug)}$\cite{FLAP} & LA+A+C & & 41.5 & 75.5 & 86.0 &-& 53.0 & 84.1 & 92.6 &-& 20.3 & 46.5 & 58.8 &-& 25.5 & 53.4 & \textbf{67.9} &- \\
GLAP\cite{Dinkel2025GLAP} & LS+WC+A+C &  & 41.7	&-& 86.1 &-& 54.4 &-& 91.1 &-& 19.4 &-& 58.3 &-& 21.8 &-& 61.5 &- \\
\midrule

\multicolumn{6}{l}{\textit{(Proposed model)}}  \\
M2D-CLAP$_\text{2025}$ & WC+A+C & \checkmark & 41.96 &\textbf{77.07}&\textbf{88.57}&\textbf{56.74}&\textbf{59.25}& {84.74} &\textbf{92.79}&\textbf{43.24}& {20.08}& {45.99}& {59.48}&\textbf{31.21}& 24.98 & 51.67 & 64.59 & 17.35 \\
\addlinespace[0.05cm]

\multicolumn{6}{l}{\textit{(Ablations)}}  \\
M2D-CLAP$^\text{ stage1}_\text{2025}$ & LS+WC & & 27.34 & 58.14 & 71.45 & 40.14 & 36.05 & 65.31 & 77.43 & 23.95 & 15.92 & 37.88 & 49.26 & 25.37 & 18.09 & 39.90 & 52.54 & 12.39 \\
\addlinespace[0.05cm]
M2D-CLAP$^\text{ stage2}_\text{2025}$ & LS+WC+A+C & & 37.24 & 72.52 & 84.54 & 52.00 & 52.56 & 82.34 & 90.60 & 38.74 & 18.09 & 42.53 & 55.52 & 28.51 & 24.88 & 48.13 & 60.48 & 15.53 \\
\addlinespace[0.05cm]
M2D-CLAP$^\text{ stage2.1}_\text{2025}$ & WC+A+C &  & 40.54 & 74.52 & 86.92 & 55.09 & 57.99 & 83.91 & 91.43 & 41.72 & 19.43 & 45.84 & 58.64 & 30.55 & 26.32 & 51.20 & 65.07 & \textbf{17.65} \\
\addlinespace[0.05cm]

\multicolumn{15}{l}{\textit{(Ablations using a non-LLM text encoder. For the training data, M2D-CLAP$^\text{ stage1}_\text{2024}$\cite{niizumi24M2D-CLAP} used AudioSet only while +WC used AudioSet and WavCaps.)}}  \\
M2D-CLAP$^\text{ stage1}_\text{2024}$\cite{niizumi24M2D-CLAP} & LS &  & 24.97 & 58.58 & 72.50 & 39.00 & 25.50 & 58.93 & 76.07 & 18.59 & 10.24 & 26.49 & 37.07 & 17.33 & 11.00 & 25.93 & 35.31 & 7.20 \\
\hspace{25pt} +WC & LS+WC & & 26.06 & 57.89 & 72.64 & 39.79 & 28.63 & 61.65 & 75.97 & 19.69 & 14.81 & 35.44 & 47.00 & 23.78 & 15.41 & 33.78 & 47.27 & 10.42 \\

\bottomrule\\
\end{tabular}
}
\vspace{-15pt}
\end{table*}

\subsubsection{Audio-Text Retrieval} \label{sec:exp-clap-atr}
The audio-text retrieval task evaluates CLAP features and involves searching for text from audio or audio from text, so unlike zero-shot classification, it requires fine-grained comparisons to find pairs at the data instance level\cite{mei22OnMetricLearning}. Therefore, task performance improves as the features more accurately align the audio and text.

\paragraph{Experimental setup}
As in previous studies\cite{Mei2023WavCaps}\cite{mei22OnMetricLearning}, we evaluated the CLAP features using the test split samples of AudioCaps\cite{kim2019audiocaps} and Clotho v2\cite{icassp20Clotho}.
In addition to the standard metrics recall at rank k (R@k), R@1, R@5, and R@10, we also report mAP@10, the primary metric for the DCASE 2024 Challenge\footnote{\scriptsize{\url{https://dcase.community/challenge2024/task-language-based-audio-retrieval}}} Task 8.

\paragraph{Results: Audio-text retrieval}
Table \ref{tab:exp-atr:results} shows that M2D-CLAP$_\text{2025}$ achieves top performance, and the ablation variants improve the performance as stages go by.
The first-stage model, M2D-CLAP$^\text{ stage1}_\text{2025}$, exhibits a large gap from the previous SOTA results, such as HTSAT-BERT-FT.
The second-stage pre-training significantly improves the performance, e.g., from 27.34 of the first stage to 37.24 of the second stage for AudioCaps Text-to-Audio R@1. This indicates that pre-training the text encoder is crucial for a task that requires finer alignment between audio and text, unlike the zero-shot classification, where the first-stage model achieved comparable results.
The additional stages 1.1 (AudioSet fine-tuning) and 2.1 (WavCaps fine-tuning) bring the final performance to the top level, and M2D-CLAP$_\text{2025}$ even outperforms top performers HTSAT-BERT-FT, Cacophony, and FLAP in AudioCaps.

M2D-CLAP$^\text{ stage1}_\text{2025}$ outperforms both M2D-CLAP$^\text{ stage1}_\text{2024}$ and M2D-CLAP$^\text{ stage1}_\text{2024}$+WC. Notably, the latter is pre-trained on the same dataset as M2D-CLAP$^\text{ stage1}_\text{2025}$, ensuring that the only differences lie in the text encoder and audio projector.
M2D-CLAP$^\text{ stage1}_\text{2025}$ improves performance on most metrics, e.g., AudioCaps Audio-to-Text R@1 performance of 36.05 compared to M2D-CLAP$^\text{ stage1}_\text{2024}$+WC's 28.63. The results confirm the effectiveness of using an LLM-based text (sentence) encoder and a transformer-based audio projector.

Table \ref{tab:exp-atr:results} indicates that the combination of datasets significantly impacts performance. The \textit{“Final CLAP training dataset”} column lists the datasets used in the model's last training just before the evaluation.
LAION-CLAP, CLAP$_{2023}$, and M2D-CLAP$^\text{ stage1\&2}_\text{2025}$, which used a large-scale dataset (denoted as "LS" in the table), perform poorly compared to those without one. Their AudioCaps Text-to-Audio R@1 performance is about 35 to 37, which is inferior to the 39.7 of HTSAT-BERT-PT without using large-scale datasets, for example. These results indicate that a limited combination of WavCaps, AudioCaps, and Clotho is advantageous for evaluation in the benchmark.
As an exception, Cacophony, which developed its data curation strategy, performs well with large-scale datasets.

In summary, the experiments validated the effectiveness of M2D-CLAP and demonstrated that the staged training process enables the model to achieve fine-grained alignment between paired features.

\begin{table*}[tb!]
%\vspace{-5pt}
\caption{Audio Captioning: CLAP audio feature comparisons on the EnCLAP codebase with 95\% CI.}
\label{tab:exp-audio-captioning:results}
\vspace{-5pt}
\centering
\resizebox{0.95\textwidth}{!}{%
\begin{tabular}{lcllllllllll}
\toprule
 & \multirow{3}{*}{\shortstack[c]{Audio enc.\\AudioSet\\fine-tuning}} & \multicolumn{5}{c}{AudioCaps (subset of AudioSet)} & \multicolumn{5}{c}{Clotho (out of AudioSet)} \\
 \cmidrule(lr){3-7} \cmidrule(lr){8-12}
CLAP model &  & METEOR & CIDEr & SPICE & SPIDEr & FENSE & METEOR & CIDEr & SPICE & SPIDEr & FENSE \\
\addlinespace[0.1cm]
\midrule
\multicolumn{6}{l}{\textit{(Baselines reproduced on the same codebase)}}  \\
LAION-CLAP\cite{LAION-CLAP} & \checkmark &23.8{\fontsize{6pt}{6pt} \selectfont $\pm$0.2} & 71.7{\fontsize{6pt}{6pt} \selectfont $\pm$0.9} & \textbf{17.6{\fontsize{6pt}{6pt} \selectfont $\pm$0.2}} & 44.6{\fontsize{6pt}{6pt} \selectfont $\pm$0.5} &\textbf{64.0{\fontsize{6pt}{6pt} \selectfont $\pm$0.3}}& 17.4{\fontsize{6pt}{6pt} \selectfont $\pm$0.1} & 41.4{\fontsize{6pt}{6pt} \selectfont $\pm$0.3} & 12.1{\fontsize{6pt}{6pt} \selectfont $\pm$0.1} & 26.8{\fontsize{6pt}{6pt} \selectfont $\pm$0.2} & 48.6{\fontsize{6pt}{6pt} \selectfont $\pm$0.2} \\
CLAP$_{2023}$\cite{CLAP2023} & \checkmark & 22.4{\fontsize{6pt}{6pt} \selectfont $\pm$0.5} & 65.2{\fontsize{6pt}{6pt} \selectfont $\pm$2.1} & 16.3{\fontsize{6pt}{6pt} \selectfont $\pm$0.3} & 40.7{\fontsize{6pt}{6pt} \selectfont $\pm$1.2} & 60.5{\fontsize{6pt}{6pt} \selectfont $\pm$0.8} & 17.5{\fontsize{6pt}{6pt} \selectfont $\pm$0.1} & 42.5{\fontsize{6pt}{6pt} \selectfont $\pm$0.5} & 12.3{\fontsize{6pt}{6pt} \selectfont $\pm$0.1} & 27.4{\fontsize{6pt}{6pt} \selectfont $\pm$0.3} & 49.2{\fontsize{6pt}{6pt} \selectfont $\pm$0.2} \\

\multicolumn{6}{l}{\textit{(Proposed model)}}  \\
M2D-CLAP$_\text{2025}$ & \checkmark &\textbf{24.3{\fontsize{6pt}{6pt} \selectfont $\pm$0.4}}&\textbf{72.4{\fontsize{6pt}{6pt} \selectfont $\pm$1.8}}&\textbf{17.6{\fontsize{6pt}{6pt} \selectfont $\pm$0.3}}&\textbf{45.0{\fontsize{6pt}{6pt} \selectfont $\pm$1.0}}& 63.9{\fontsize{6pt}{6pt} \selectfont $\pm$0.4} & 17.7{\fontsize{6pt}{6pt} \selectfont $\pm$0.1} & 43.1{\fontsize{6pt}{6pt} \selectfont $\pm$0.7} &\textbf{12.5{\fontsize{6pt}{6pt} \selectfont $\pm$0.2}}& 27.8{\fontsize{6pt}{6pt} \selectfont $\pm$0.4} & 49.6{\fontsize{6pt}{6pt} \selectfont $\pm$0.4} \\
\addlinespace[0.05cm]

\multicolumn{6}{l}{\textit{(Ablations)}}  \\

M2D-CLAP$^\text{ stage1}_\text{2025}$ & -& 23.3{\fontsize{6pt}{6pt} \selectfont $\pm$0.3} & 68.3{\fontsize{6pt}{6pt} \selectfont $\pm$1.5} & 16.9{\fontsize{6pt}{6pt} \selectfont $\pm$0.2} & 42.6{\fontsize{6pt}{6pt} \selectfont $\pm$0.8} & 61.9{\fontsize{6pt}{6pt} \selectfont $\pm$0.5} & 17.4{\fontsize{6pt}{6pt} \selectfont $\pm$0.1} & 41.2{\fontsize{6pt}{6pt} \selectfont $\pm$0.3} & 12.2{\fontsize{6pt}{6pt} \selectfont $\pm$0.1} & 26.7{\fontsize{6pt}{6pt} \selectfont $\pm$0.2} & 48.9{\fontsize{6pt}{6pt} \selectfont $\pm$0.2} \\
\addlinespace[0.05cm]
M2D-CLAP$^\text{ stage2}_\text{2025}$ & -& 23.7{\fontsize{6pt}{6pt} \selectfont $\pm$0.3} & 69.6{\fontsize{6pt}{6pt} \selectfont $\pm$1.3} & 17.2{\fontsize{6pt}{6pt} \selectfont $\pm$0.3} & 43.4{\fontsize{6pt}{6pt} \selectfont $\pm$0.7} & 62.5{\fontsize{6pt}{6pt} \selectfont $\pm$0.4} & \textbf{17.8{\fontsize{6pt}{6pt} \selectfont $\pm$0.1}} & \textbf{43.5{\fontsize{6pt}{6pt} \selectfont $\pm$0.4}} & 12.4{\fontsize{6pt}{6pt} \selectfont $\pm$0.1} & \textbf{28.0{\fontsize{6pt}{6pt} \selectfont $\pm$0.2}} & \textbf{49.9{\fontsize{6pt}{6pt} \selectfont $\pm$0.2}} \\
\addlinespace[0.05cm]
M2D-CLAP$^\text{ stage2.1}_\text{2025}$ & -& 23.6{\fontsize{6pt}{6pt} \selectfont $\pm$0.3} & 69.4{\fontsize{6pt}{6pt} \selectfont $\pm$1.4} & 17.1{\fontsize{6pt}{6pt} \selectfont $\pm$0.3} & 43.2{\fontsize{6pt}{6pt} \selectfont $\pm$0.8} & 62.3{\fontsize{6pt}{6pt} \selectfont $\pm$0.5} & \textbf{17.8{\fontsize{6pt}{6pt} \selectfont $\pm$0.1}} & 43.2{\fontsize{6pt}{6pt} \selectfont $\pm$0.4} & \textbf{12.5{\fontsize{6pt}{6pt} \selectfont $\pm$0.1}} & 27.8{\fontsize{6pt}{6pt} \selectfont $\pm$0.3} & 49.7{\fontsize{6pt}{6pt} \selectfont $\pm$0.2} \\
\addlinespace[0.05cm]

\midrule

\multicolumn{6}{l}{\textit{(Reference results from non-EnCLAP models)}}  \\
CNN14-BART\cite{Mei2023WavCaps} & \checkmark & {\color{EMgray}24.7} &{\color{EMgray}75.6} &{\color{EMgray}17.9} &{\color{EMgray}46.8} & - & {\color{EMgray}18.5} & {\color{EMgray}48.8} & {\color{EMgray}13.3} & {\color{EMgray}31.0} &- \\
HTSAT-BART\cite{Mei2023WavCaps} & \checkmark & {\color{EMgray}25.0} &{\color{EMgray}78.7} &{\color{EMgray}18.2} &{\color{EMgray}48.5} & - & {\color{EMgray}18.4} & {\color{EMgray}46.2} & {\color{EMgray}13.3} & {\color{EMgray}29.7} &- \\
CLAP$_{2023}$\cite{CLAP2023} & \checkmark &-&-&-&{\color{EMgray}45.5}& - &-&-&-&{\color{EMgray}27.1}& -\\
Cacophony\cite{Zhu24Cacophony} &  & {\color{EMgray}23.6} & {\color{EMgray}72.8} & {\color{EMgray}16.8} & {\color{EMgray}44.8} & - & {\color{EMgray}15.3} & {\color{EMgray}34.2} & {\color{EMgray}10.6} & {\color{EMgray}22.4} & -\\
\bottomrule\\
\end{tabular}
}
\vspace{-10pt}
\end{table*}

\subsubsection{Audio Captioning} \label{sec:exp-clap-captioning}
We evaluated the effectiveness of the CLAP audio feature on the audio captioning task using a baseline system that relies on a CLAP model to extract an audio semantic feature to generate audio captions.
In the baseline system, the CLAP audio features provide pivotal information to represent the input audio semantics; thus, the better these features represent the audio semantics, the better the output captions.

\paragraph{Experimental setup}
We used the official codebase of a SOTA method, EnCLAP\footnote{\scriptsize{\url{https://github.com/jaeyeonkim99/EnCLAP}}}\cite{EnCLAP}, and replaced the CLAP used in EnCLAP, thereby evaluating only the differences in CLAP under a unified environment. EnCLAP uses CLAP audio features to represent the semantics of the audio clip and EnCodec\cite{encodec} tokens as supplemental time series information to generate captions using BART\cite{BART}. We replaced the CLAP only in the codebase during the evaluation.

We conducted evaluations on AudioCaps and Clotho and used a learning rate of $6.0\times 10^{-6}$ for AudioCaps and $4.0\times 10^{-6}$ for Clotho. We used the same settings for other hyperparameters as in EnCLAP, such as training epochs of 15.
We tested the top five checkpoints of SPIDEr validation performance on the test set and averaged the results to obtain the final result. The evaluation metrics are METEOR\cite{METEOR}, CIDEr\cite{CIDEr}, SPICE\cite{SPICE}, SPIDEr\cite{SPIDEr}, and FENSE\cite{FENSE}.

\paragraph{Results: Audio captioning}
Table \ref{tab:exp-audio-captioning:results} shows that M2D-CLAP$_\text{2025}$ archives top performance, which is comparable to LAION-CLAP on AudioCaps and better than LAION-CLAP on Clotho. We also evaluated CLAP$_{2023}$; however, it underperformed LAION-CLAP. Note that the experiments used the same EnCLAP codebase; the only difference is the CLAP audio feature used by EnCLAP.

The AudioCaps results of intermediate stages, M2D-CLAP$^{\text{ stage}\{1, 2, 2.1\}}_\text{2025}$, underperform those of LAION-CLAP, indicating that fine-tuning the audio encoder to AudioSet is vital for good performance with AudioCaps.
On the other hand, the Clotho results suggest that fine-tuning to AudioSet does not offer general-purpose performance; all M2D-CLAP variants outperform LAION-CLAP.

Notably, M2D-CLAP$^\text{ stage2}_\text{2025}$ performs slightly better than M2D-CLAP$_\text{2025}$ on Clotho, demonstrating that it learns generalizable CLAP features without the need for fine-tuning to AudioSet and WavCaps.

In summary, the experimental results confirm that M2D-CLAP produces CLAP features that also perform well in audio captioning.

\subsection{M2D-CLAP Analysis} \label{sec:exp-ablations}

\begin{figure}[tbp]
  \vspace{-10pt}
  \centering
  \includegraphics[width=1.0\columnwidth]{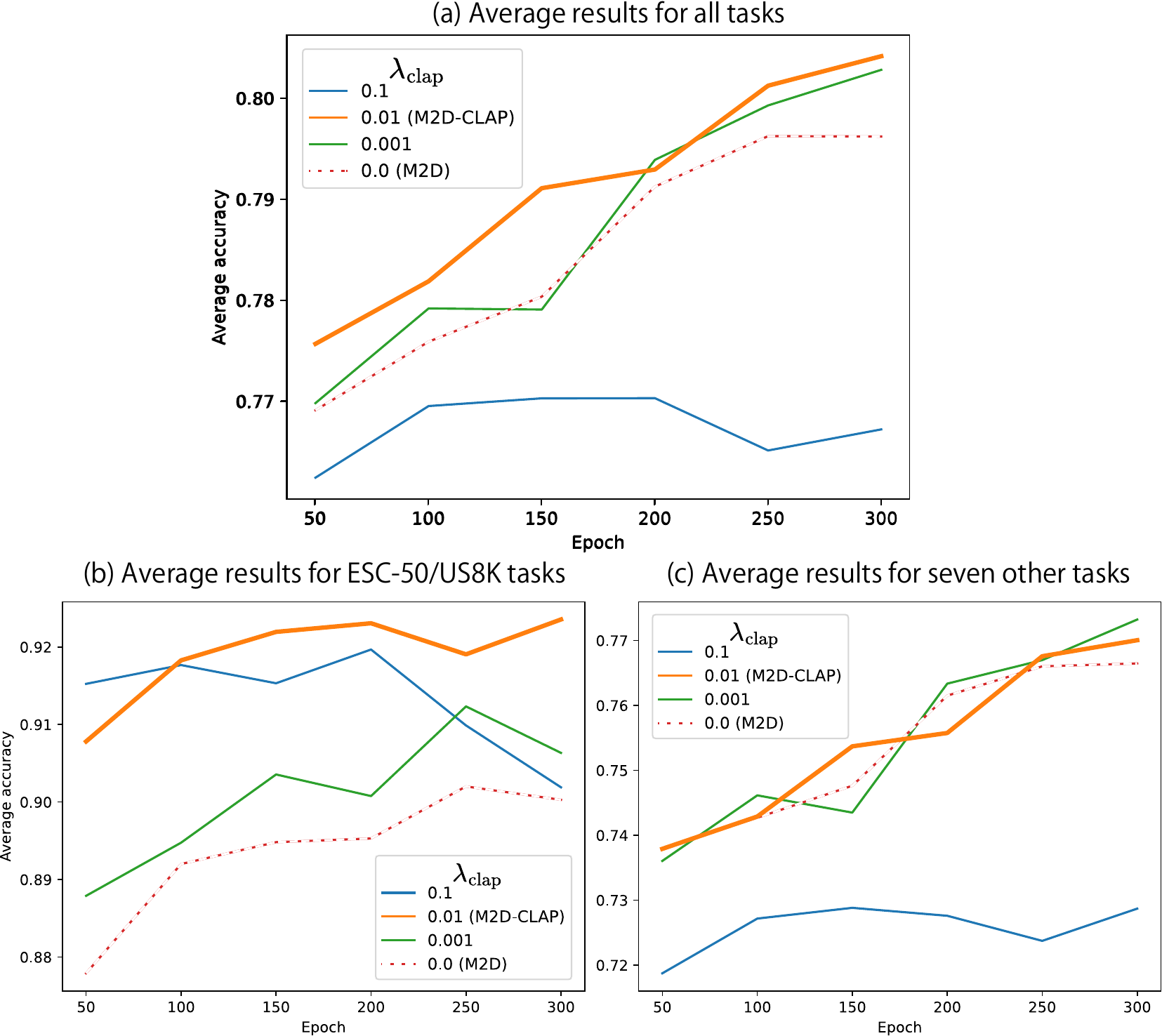}
  \vspace{-10pt}
  \caption{CLAP loss weight ablations: Average linear evaluation performance comparisons show that the large CLAP loss weight $\lambda_\text{clap}$ corrupts the training while the small $\lambda_\text{clap}$ weakens the performance.}
  \label{fig:C-abl-CLAP-loss-weight}
  \vspace{-10pt}
\end{figure}

\begin{figure}[tbp]
  \vspace{-10pt}
  \centering
  \includegraphics[width=1.0\columnwidth]{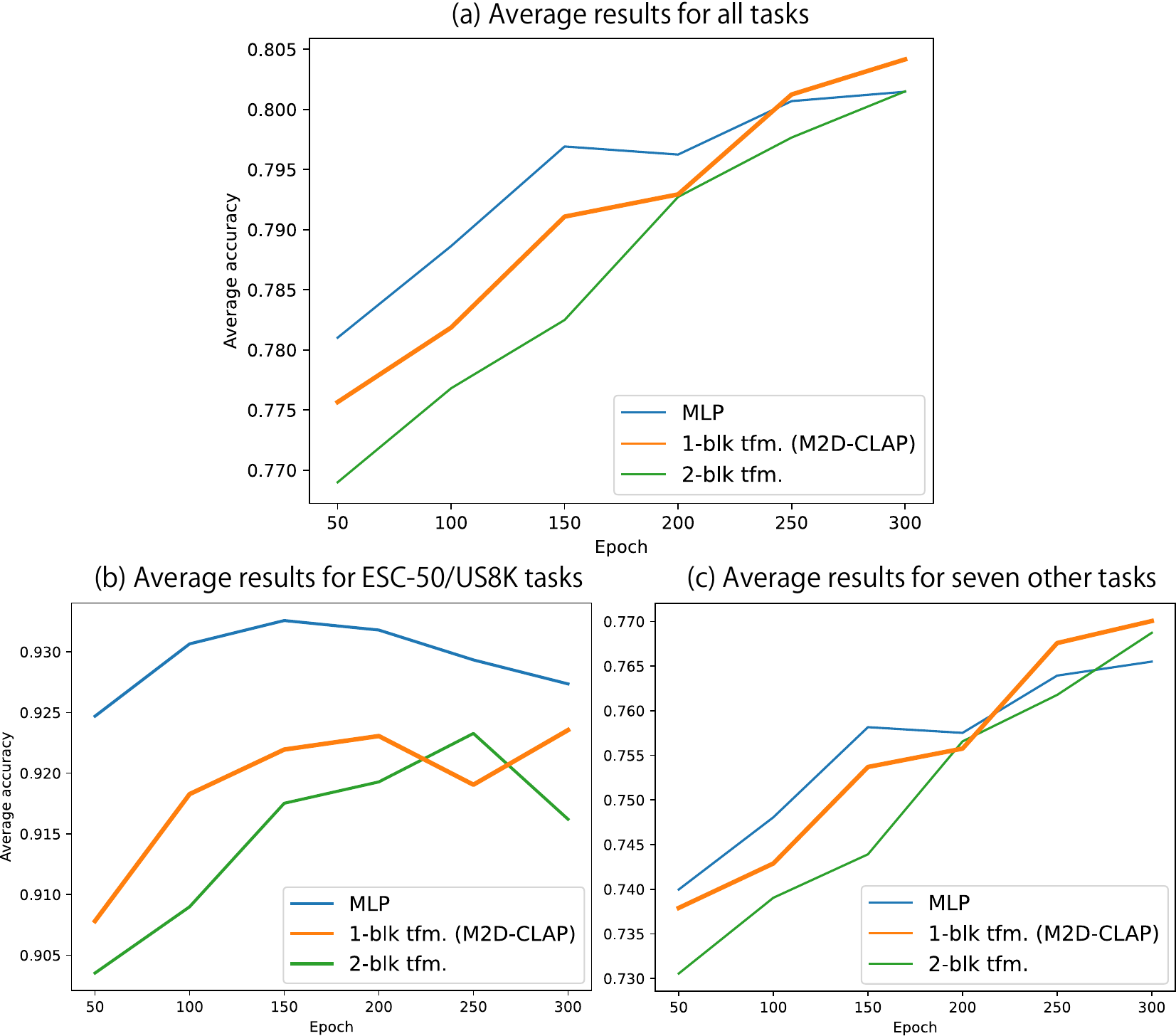}
  \vspace{-10pt}
  \caption{Audio projector ablations: Average linear evaluation performance comparison shows that the one-block projector keeps increasing performance, whereas others degrade ESC-50/US8K results at the end.}
  \label{fig:C-abl-audio-proj}
  \vspace{-10pt}
\end{figure}

\subsubsection{Taming the CLAP objective in the first stage} \label{sec:exp-abl-clap-in-1st}
While training with the CLAP objective provides a powerful training signal for the semantic features, it is challenging to balance it with M2D masked prediction learning, which trains effective general-purpose representations.
We investigated the impact of CLAP loss weight adjustment and audio projector design on the performance of the learned representations.

\paragraph{CLAP loss weight ablations} %\label{sec:exp-abl-clap-loss}
We investigated the effect of CLAP loss weight $\lambda_\text{clap}$ in Eq. \eqref{eq:eq-m2d-clap-loss}, which directly adjusts the effect of the CLAP objective. Fig. \ref{fig:C-abl-CLAP-loss-weight} shows the average results of linear evaluation on general tasks across training epochs for (a)~all tasks, (b)~ESC-50/US8K tasks for which the CLAP objective is effective, and (c)~all other tasks when the weight $\lambda_\text{clap}$ is varied.

With a weight of 0.1, (b) ESC-50/US8K performance matures quickly but degrades later, while (c) other tasks show no improvement. At 0.001, (b) performance improves slowly, indicating the weight is too low. In contrast, 0.01 allows (a) all tasks to mature steadily during training.
For reference, at weight 0.0 (i.e., M2D), (b) is low due to lack of caption supervision, (c) remains high, and (a) is slightly lower overall than M2D-CLAP.

The rapid maturation of ESC-50/US8K performance with higher CLAP weights mirrors the early maturation of CLAP training reported in prior work\cite{CLAP2023}. These results suggest that suppressing the effect of CLAP while combining it with SSL is key for effective pre-training.

\paragraph{Audio projector ablations} \label{sec:exp-abl-audio-proj}
The audio projector, which aggregates token-level audio features into a single feature for CLAP loss computation, impacts training effectiveness.
Fig. \ref{fig:C-abl-audio-proj} compares average linear evaluation results across epochs for different audio projectors: the one-block transformer in M2D-CLAP, a two-block transformer (with one additional block), and a one-hidden-layer MLP used in the previous M2D-CLAP\cite{niizumi24M2D-CLAP}.
For the MLP, we input the average of all audio encoder outputs.

In Fig.~\ref{fig:C-abl-audio-proj}(a), the MLP performs best up to 200 epochs but stalls afterward. In the final 300 epochs, M2D-CLAP's one-block transformer surpasses the others.
The MLP's trend (Fig.~\ref{fig:C-abl-audio-proj}(b)) resembles ESC-50/US8K results under a high CLAP loss weight ($\lambda_\text{clap}=0.1$) in Fig.~\ref{fig:C-abl-CLAP-loss-weight}(b), suggesting that a simple MLP increases CLAP’s influence during training and leads to early maturation, in line with previous findings\cite{CLAP2023}.

The two-block transformer consistently underperforms the one-block version, indicating that moderate network capacity is important for effective pre-training.

\subsubsection{Text encoder ablations} \label{sec:exp-abl-semantic-feature}
We assessed the performance impact of the type of text encoder used in the first stage. We compared the LLM-based model NV-Embed-v2\cite{NV-Embed}\cite{NV-Embed2} (7.85B parameters) used in M2D-CLAP against the GTE-Base\cite{li2023GTE} (109M parameters) used in the previous M2D-CLAP (M2D-CLAP$^\text{ stage1}_\text{2024}$) and GTE-Qwen2-7B-instruct\footnote{\url{https://huggingface.co/Alibaba-NLP/gte-Qwen2-7B-instruct}} (7.61B parameters)\cite{li2023GTE}, which ranked higher than NV-Embed-v2 on the MTLB leaderboard. The experiments replaced the text encoder in M2D-CLAP.
We also assessed the performance of the second-stage text encoder when using GTE-Base instead of BERT.

The linear evaluation results on the general audio tasks in Table \ref{tab:exp-LE:abl-NorA} and the audio-text retrieval of the CLAP features in Table \ref{tab:exp-atr:abl-NorA} show that using NV-Embed-v2's LLM-based sentence features achieves the best performance with both features.
With a non-LLM-based GTE-Base, M2D-CLAP learns competitive audio features while underperforming second-stage CLAP features compared to LLM-based encoders. 
The GTE-Qwen2-7B-instruct's results are comparable to (while slightly worse than) NV-Embed-v2's, confirming the effectiveness of employing LLM-based text encoders.

\begin{table}[tb!]
%\vspace{-5pt}
\caption{Text encoder ablations: Linear evaluation comparison on general audio tasks using audio feature.}
\label{tab:exp-LE:abl-NorA}
\vspace{-5pt}
\resizebox{1.0\columnwidth}{!}{%
\begin{tabular}{lllllllllll}
\toprule
Text encoder & ESC & US8K & SPC & VC1 & VF & CRD & GTZ & NS & SUR & Avg. \\
\midrule
NV-Embed-v2 & 95.2 &\textbf{89.5}& {95.6} &\textbf{68.6}&\textbf{98.1}& 73.4 &\textbf{86.6}& \textbf{76.5}& {42.2}&\textbf{80.6}\\
GTE-Qwen2-7B &\textbf{95.4}& 89.1 &\textbf{95.9}& 67.7 & 98.1 & 73.6 & 85.2 & 75.0 &\textbf{42.4}& 80.3 \\
GTE-Base & {95.3}& 89.2 &{95.6}& 66.2 & 97.7 &\textbf{74.7}& 84.3 & 75.2 &  {42.2} & 80.0 \\
\bottomrule\\
\end{tabular}\\
}
\vspace{-5pt}
\end{table}

\begin{table}[tb!]
%\vspace{-5pt}
\caption{Text encoder ablations: Audio-text retrieval comparison using CLAP feature.}
\label{tab:exp-atr:abl-NorA}
\vspace{-5pt}
\centering
\resizebox{0.95\columnwidth}{!}{%
\begin{tabular}{lcccccccc}
\toprule
 \multirow{4}{*}{\shortstack[c]{Text encoder\\used in the\\\textbf{first stage}}} & \multicolumn{4}{c}{AudioCaps} & \multicolumn{4}{c}{Clotho} \\
 \cmidrule(lr){2-5} \cmidrule(lr){6-9}
 & \multicolumn{2}{c}{T-to-A} & \multicolumn{2}{c}{A-to-T} & \multicolumn{2}{c}{T-to-A} & \multicolumn{2}{c}{A-to-T} \\
 & R@1 & \scriptsize{\shortstack[c]{mAP\\@10}}  & R@1 & \scriptsize{\shortstack[c]{mAP\\@10}} & R@1 & \scriptsize{\shortstack[c]{mAP\\@10}} & R@1 & \scriptsize{\shortstack[c]{mAP\\@10}}\\
\midrule
\multicolumn{6}{l}{\textit{(The first-stage pre-training)}}  \\
NV-Embed-v2 &  \textbf{27.34}& 40.14 &\textbf{36.05}&\textbf{23.95}&\textbf{15.92}&\textbf{25.37}&\textbf{18.09}&\textbf{12.39} \\
GTE-Qwen2-7B & 21.88 & 34.35 & 31.45 & 20.05 & 14.37 & 22.88 & 16.08 & 10.31 \\
GTE-Base & 26.96 &\textbf{40.45}& 32.18 & 21.95 & 14.93 & 23.85 & 17.03 & 11.34 \\

\midrule
\multicolumn{9}{l}{\textit{(The second-stage pre-training using BERT as a text encoder)}}  \\
NV-Embed-v2 &\textbf{37.24}&\textbf{52.00}&\textbf{52.56}&\textbf{38.74}&\textbf{18.09}&\textbf{28.51}&\textbf{24.88}&\textbf{15.53}\\
GTE-Qwen2-7B & 36.78 & 51.64 & 51.41 & 38.08 & 17.86 & 28.36 & 23.25 & 15.16 \\
GTE-Base & 34.04 & 48.57 & 50.37 & 37.29 & 16.17 & 26.13 & 19.23 & 12.52 \\

\midrule
\multicolumn{9}{l}{\textit{(Using text encoder NV-Embed-v2 at the first stage and GTE-Base at the second stage)}}  \\
NVEv2$\rightarrow$GTE &\textbf{37.28}& 51.77 & 48.17 & 37.03 &\textbf{18.24}&\textbf{28.79}& 24.59 & {15.40}\\
 & \multicolumn{8}{l}{\scriptsize{{Bold results are better than the NV-Embed-v2 (NVEv2$\rightarrow$BERT) results above.}}}  \\
\bottomrule\\
\end{tabular}
}
\vspace{-10pt}
\end{table}

Notably, the performance boost in the second-stage CLAP features indicates that the semantic knowledge accumulated in the audio features effectively contributes to learn useful CLAP features.
The second-stage results of LLM-based NV-Embed-v2 and GTE-Qwen2-7B-instruct are 37.24 and 36.78 for R@1 of AudioCaps T-to-A, while the non-LLM-based GTE-Base's result is 34.04, for example.

The results demonstrate that powerful generalist LLM-based sentence embedding models with SOTA performance in the NLP domain are also useful for learning audio and CLAP features. This finding also suggests that utilizing generalist embedding models in other audio-language tasks would be beneficial, and we leave the possibilities to future work.

The NVEv2$\rightarrow$GTE results confirm that using a generalist sentence embedding model (GTE-Base) instead of the typically used BERT for the second stage does not improve the final CLAP feature performance. While it performs slightly better than NV-Embed-v2 (NV-Embed-v2$\rightarrow$BERT) for T-to-A, it degrades more for the AudioCaps A-to-T.

\subsubsection{Training design ablations} \label{sec:exp-abl-training-design}
We conducted the following ablations focusing on differences from conventional CLAP in the M2D-CLAP training process:

\begin{enumerate}[(i)]
   \item Training audio encoder in the second stage:
   While M2D-CLAP freezes the audio encoder in the second stage, an alternative is to pre-train both audio and text encoders, as in \cite{Zhu24Cacophony}. 
   This ablation adopts that approach, while reducing the batch size to 1024 from 2048 due to limited computational resources.
   \item Using text projector only:
   This ablation removed the audio projector and added a text projector on top of the text encoder, aiming to learn an audio feature directly in the semantic space that serves both conventional and CLAP audio features.
We used a one-hidden-layer MLP as the text projector.
   \item Adding text projector:
   This ablation added a text projector on top of the text encoder, making M2D-CLAP follow a typical CLAP configuration with both audio and text projectors.
\end{enumerate}

%\vspace{0.2cm}
Tables \ref{tab:exp-LE:abl-audio-trn-at-2nd} and \ref{tab:exp-atr:abl-audio-trn-at-2nd} present the evaluation results for audio and CLAP features, respectively, with key observations summarized below:

\textit{(i) Training audio encoder in the second stage:}
For audio features, the ablation (Stage 2/TA) improves performance on ESC-50 (ESC) and US8K, but clearly degrades it on speech tasks. This trend, similar to that of the final M2D-CLAP$_\text{2025}$, suggests an influence of caption supervision.

For CLAP features, Stage 2/TA lowers A-to-T performance in both retrieval tasks. Unlike previous studies, factors such as pre-training the audio encoder under M2D and CLAP objectives, omitting the text projector, or using a non-MLP audio projector may have slightly hindered fine-grained audio-to-text alignment. Additionally, the reduced batch size may have negatively impacted results.
   
\textit{(ii) Using text projector only:}
Unlike other ablations, the ablations (Stages 1, 2/TP only) degrade performance across all results for both features. Thus, these results highlight the challenge of bridging conventional and CLAP audio features.

\textit{(iii) Adding text projector:}
The ablation (Stages 1/+TP) causes minimal changes in audio feature performance, while Stages 2/+TP clearly degrades CLAP feature performance. Mapping text features through a projector appears to hinder overall audio-text alignment in CLAP features.

\vspace{0.2cm}
In summary, unlike previous work such as Cacophony\cite{Zhu24Cacophony}, continuing to train the audio encoder in Stage 2 is ineffective, and incorporating a text projector yields no clear advantage for M2D-CLAP.

\begin{table}[tb!]
%\vspace{-5pt}
\caption{Training design ablations for audio features: Linear evaluation comparison. Bold improves over Stage 1; others degrade.}
\label{tab:exp-LE:abl-audio-trn-at-2nd}
\vspace{-5pt}
\resizebox{1.0\columnwidth}{!}{%
\begin{tabular}{lllllllllll}
\toprule
Model & ESC & US8K & SPC & VC1 & VF & CRD & GTZ & NS & SUR & Avg. \\
\midrule
 Stage 1 & 95.2 & 89.5 & {95.6}& {68.6}& {98.1}& {73.4}&  {86.6} & 76.5 & {42.2}& {80.6}\\
\midrule
\multicolumn{10}{l}{\textit{(i) Training audio encoder in the second stage.}}  \\
Stage 2/TA & \textbf{97.0} &\textbf{90.1}& 93.3 & 60.4 & 96.7 & 71.1 &86.3& 76.1 & 42.0 & 79.0 \\
\addlinespace[0.05cm]
\multicolumn{10}{l}{\textit{(ii) Using text projector only.}}  \\
Stage 1/TP only & 94.9 & 88.0 & 94.0 & 60.5 & 97.5 & 71.6 & 83.7 & 76.0 & 41.9 & 78.7 \\
\addlinespace[0.05cm]
\multicolumn{10}{l}{\textit{(iii) Adding text projector, as in previous CLAP models.}}  \\
Stage1/+TP & \textbf{96.1} & 88.9 &\textbf{95.6}& 66.2 & 97.8 & 72.7 & 84.7 &\textbf{77.6}&\textbf{42.3}& 80.2 \\
\midrule
\addlinespace[0.1cm]
M2D-CLAP$_\text{2025}$ & {97.9}& 89.7 & 94.8 & 56.3 & 97.1 & 69.7 & 86.3 & {76.7}& 41.1 & 78.8 \\
\bottomrule\\
\end{tabular}\\
}
\vspace{-5pt}
\end{table}

\begin{table}[tb!]
%\vspace{-5pt}
\caption{Training design ablations for CLAP features: Audio-text retrieval comparison. Bold improves over Stage 2; others degrade.}
\label{tab:exp-atr:abl-audio-trn-at-2nd}
\vspace{-5pt}
\centering
\resizebox{\columnwidth}{!}{%
\begin{tabular}{lcccccccc}
\toprule
  & \multicolumn{4}{c}{AudioCaps} & \multicolumn{4}{c}{Clotho} \\
 \cmidrule(lr){2-5} \cmidrule(lr){6-9}
Model & \multicolumn{2}{c}{T-to-A} & \multicolumn{2}{c}{A-to-T} & \multicolumn{2}{c}{T-to-A} & \multicolumn{2}{c}{A-to-T} \\
 & R@1 & \scriptsize{\shortstack[c]{mAP\\@10}}  & R@1 & \scriptsize{\shortstack[c]{mAP\\@10}} & R@1 & \scriptsize{\shortstack[c]{mAP\\@10}} & R@1 & \scriptsize{\shortstack[c]{mAP\\@10}}\\
\midrule
Stage 2 & 37.24 & 52.00 & 52.56 & 38.74 & 18.09 & 28.51 & {24.88}& 15.53 \\
\addlinespace[0.05cm]
\midrule
\multicolumn{9}{l}{\textit{(i) Training audio encoder in the second stage.}}  \\
Stage 2/TA & \textbf{37.81} & \textbf{52.29} & 50.78 & 37.15 & 17.76 & \textbf{28.83} & 21.44 & 14.11 \\
\addlinespace[0.05cm]
\multicolumn{9}{l}{\textit{(ii) Using text projector only.}}  \\
Stage 2/TP only & 34.04 & 48.73 & 45.98 & 34.42 & 17.07 & 27.44 & 22.11 & 14.56 \\
\addlinespace[0.05cm]
\multicolumn{9}{l}{\textit{(iii) Adding text projector, as in previous CLAP models.}}  \\
Stage 2/+TP & 36.07 & 50.52 & 48.69 & 36.93 & 17.53 & 28.09 & 22.58 & 14.72 \\
\midrule
\addlinespace[0.1cm]
M2D-CLAP$_\text{2025}$ &  {39.06}& {54.29}& {54.02}& {39.95}& {18.30}& {29.15}& 22.78 & {15.62}\\
\addlinespace[0.05cm]
\bottomrule\\
\end{tabular}
}
\vspace{-10pt}
\end{table}

\subsubsection{Visualizations} \label{sec:exp-abl-viz}

\paragraph{CLAP audio feature visualizations} \label{sec:exp-abl-feature-viz-clap}

\begin{figure}[tbp]
  \vspace{-5pt}
  \centering
  \includegraphics[width=0.9\columnwidth]{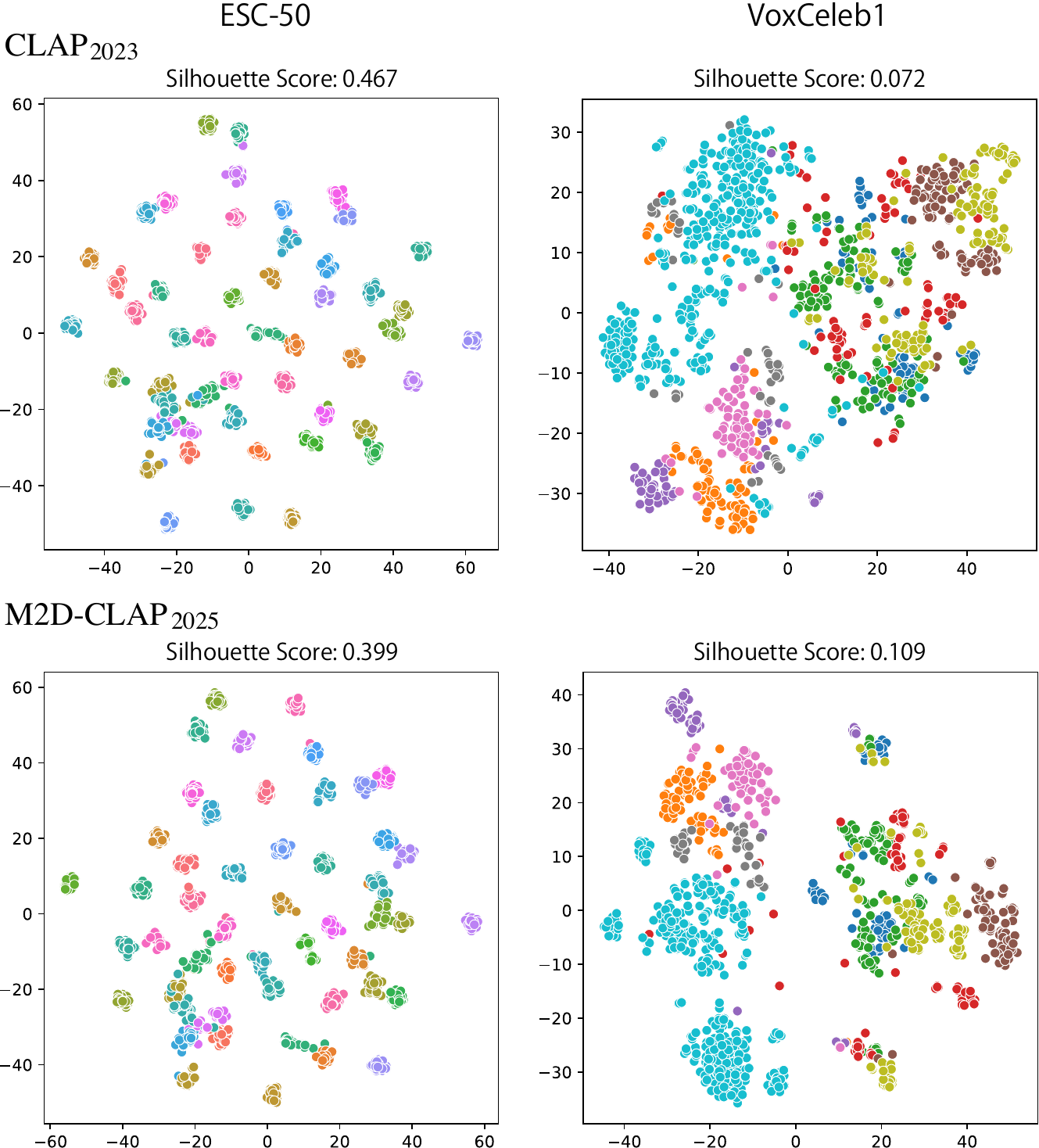}
  %\vspace{-5pt}
  \caption{Audio features visualizations. ESC-50 sample features from both models (left column) form clusters for each class.
  In contrast, the VoxCeleb1 ten-speaker sample features (right column) of CLAP$_\text{2023}$ form slightly less distinct clusters for some speakers compared to those of M2D-CLAP$_\text{2025}$.}
  \label{fig:C-viz-audio-feat}
  \vspace{-15pt}
\end{figure}

We visualize the pre-trained CLAP audio features and show their impact on two downstream tasks with an inherently different nature. We encoded ESC-50 and VoxCeleb1 sounds using M2D-CLAP$_\text{2025}$ and CLAP$_\text{2023}$ and visualized their features using t-SNE as shown in Fig. \ref{fig:C-viz-audio-feat}.
We use all 50 class samples from ESC-50 and samples of random ten speakers from VoxCeleb1.
We also computed the silhouette score to quantitatively assess the quality of clusters formed by each model’s features.

The ESC-50 results show that both model features form clear class-wise clusters with high silhouette scores, consistent with their strong performance in the linear evaluation results in Table \ref{tab:results-le} in Section \ref{sec:exp-audio-LE}.
On the other hand, the VoxCeleb1 results show that CLAP$_\text{2023}$ forms slightly less distinct clusters (silhouette score of 0.072) compared to M2D-CLAP$_\text{2025}$ (silhouette score of 0.109). These observations support the VC1 performance difference in Table \ref{tab:results-le}.

The CLAP features are advantageous for ESC-50 since it is an environmental sound classification task, and the pre-training captions contain much information indicating differences in environmental sounds. On the other hand, VoxCeleb1 is a 1251-speaker identification task, and the pre-training captions cannot contain speaker names or IDs, thus making it challenging for conventional CLAP features in the identification task. The conventional CLAP model would learn to represent speaker attributes such as adult or child described in captions, which may explain the tendency not to cluster by individuals. In contrast, M2D-CLAP features form more distinct clusters, indicating that SSL helps learn more descriptive information about sound itself, which leads to clearer clusters for individuals.

These observations demonstrate that jointly learning SSL and CLAP allows M2D-CLAP to encode both discriminative and descriptive information in CLAP audio features.

\paragraph{Audio projector attention visualizations} \label{sec:exp-abl-feature-viz-att}

\begin{figure}[tbp]
  \vspace{-5pt}
  \centering
  \includegraphics[width=1.0\columnwidth]{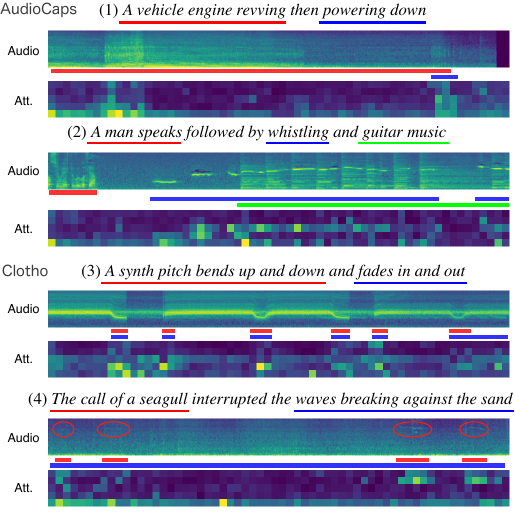}
  \vspace{-10pt}
  \caption{Audio projector attention visualizations. Examples from AudioCaps and Clotho show the caption, 10-s audio spectrogram, and attention map in the audio projector transformer. The attention map corresponds to each input audio patch, and the lighter color indicates stronger attention. We manually annotated the colored lines to show the correspondence of the events in the captions and observed event sounds in the audio sample.}
  \label{fig:C-viz-proj-att}
  \vspace{-10pt}
\end{figure}

Figure \ref{fig:C-viz-proj-att} shows the attention in the audio projector in M2D-CLAP$_\text{2025}$ for the audio samples from AudioCaps and Clotho. The attention map directly shows which patch features the final CLAP audio feature attends to because the audio projector is a single-layer transformer encoder with a single head. We picked the sounds with typical attention maps.

The first example from AudioCaps shows that the projector attends to the louder part of the revving sound and, notably, to the sound characterizing the powering down. The second shows that while the lowest frequency patches attend to speech and music, the patches at the pitch of the whistling attend to each whistling note. The third also shows that the patches at the pitch of the synth sound attend stronger to the corresponding bending up and down or fading in and out. The last example follows the same trend: the lowest frequency patches attend to the sound of waves, while the patches at the seagull calls attend to these sounds.
Overall, the observations exemplify that the projector summarizes CLAP audio features based on the patch features that characterize the audio clips.

\subsubsection{Potential and limitations} \label{sec:potential-limitation}
The application of top-performing audio features to medical (respiratory and heart) sounds in \cite{Niizumi2025EMBC} demonstrates their practical utility, suggesting broader potential for real-world problem solving. CLAP features have also been applied to diverse tasks, such as music generation\cite{MusicLDM} and foley sound synthesis\cite{LatentCLAP}. Thus, M2D-CLAP has the potential to contribute to a wide range of applications by providing both general audio and CLAP features.

On the other hand, it still has limitations in extending its applicability, particularly regarding time resolution and frequency range. It processes sounds with a temporal resolution of 160 ms and a sampling rate of 16 kHz, and adapting it to tasks with different requirements would necessitate retraining. While effectiveness has been shown even with limited frequency content, such as 4 kHz audio in \cite{Niizumi2025EMBC}, extending the approach to more diverse music tasks requiring a wider frequency range, as well as to the bioacoustic domain requiring focus on specific frequency ranges depending on the task, remains a future research direction.

The CLAP features of M2D-CLAP also have a limitation in the data coverage for general-purpose usage.
Achieving general-purpose performance, as in cutting-edge LLM, may require a similar scale of training data. To this end, further advancement of approaches such as automated generation of audio captions, as in Auto-ACD\cite{sun2023autoacd}, is necessary to address the challenge of data scalability.

\section{Conclusion}
This study pursued a general-purpose audio-language representation that provides both general-purpose audio and CLAP features to serve diverse audio applications.
For this purpose, we proposed M2D-CLAP, a novel method combining an SSL M2D and CLAP in a multi-stage pre-training and utilizing a recent LLM-based sentence encoder.
In the first stage, M2D-CLAP learns a generalizable audio feature using the rich semantic features of the LLM-based sentence embedding model, and in the second stage, it learns a CLAP feature using the learned audio feature. It further refines features by fine-tuning.
Experiments validated the effectiveness of M2D-CLAP as a general-purpose audio-language representation. Its audio features performed well, particularly in AudioSet fine-tuning and music tasks, and its CLAP features performed well in zero-shot classification, audio-text retrieval, and audio captioning. These results demonstrate the novelty of M2D-CLAP in providing both high-performing general audio and CLAP features.
We conducted ablation studies to analyze M2D-CLAP and found that a multi-stage training strategy design and managing the effect of the CLAP objective are vital for learning effective features.
We release our models to advance diverse audio studies.

% \begin{thebibliography}{1}
\bibliographystyle{IEEEtran}
\bibliography{refs}

\end{document}